\documentclass[]{jfm}

\usepackage{graphicx}
\usepackage{newtxtext}
\usepackage{newtxmath}
\usepackage{natbib}

\usepackage{commath}
\usepackage{amssymb}
\usepackage{amsmath}
\usepackage[english]{babel}
\usepackage[T1]{fontenc}
\usepackage[utf8]{inputenc}
\usepackage{todonotes}
\usepackage{tabularx}

\usepackage{hyperref}

\newcommand{\RomanNumeralCaps}[1]
\linenumbers

\usetikzlibrary{decorations.pathreplacing}
\usetikzlibrary{calc,3d,shadings}
\usetikzlibrary{arrows.meta}

\definecolor{mNGIDarkRed}{HTML}{AA2227}
\definecolor{mNGILightRed}{HTML}{EB7F6C}

\definecolor{mNGIDarkBlue}{HTML}{2c7bb6}
\definecolor{mNGILightBlue}{HTML}{abd9e9}

\renewcommand{\r}[1]{\mathrm{#1}}
\def\b#1{\mbox{\boldmath $#1$}}
\newcommand{\bs}[1]{\boldsymbol{#1}}
\newcommand{\bt}[1]{\mathsfbi{#1}}
\newcommand{\ol}[1]{\overline{#1}}

\hypersetup{
    colorlinks = true,
    urlcolor   = blue,
    citecolor  = black,
}

\graphicspath{{./graphicsmake/}}

\title{The compressible granular collapse in a fluid as a continuum: validity of a Navier-Stokes model with $\mu(J)$-$\phi(J)$-rheology}

\author{Matthias Rauter \aff{1,2,3}
  \corresp{\email{matthiar@uio.no}}}

\affiliation{\aff{1}University of Oslo, Oslo, Norway
\aff{2}Norwegian Geotechnical Institute, Oslo, Norway
\aff{3}University of Innsbruck, Innsbruck, Austria}

\begin{document}

\maketitle


\begin{abstract}
The incompressible $\mu(I)$-rheology has been used to study subaerial granular flows with remarkable success.
For subaquatic granular flows, drag between grains and the pore fluid is substantially higher and the physical behaviour is more complex.
High drag forces constrain the rearrangement of grains and dilatancy, leading to a considerable build-up of pore pressure.
Its transient and dynamic description is the key to modelling subaquatic granular flows but out of the scope of incompressible models.
In this work, we advance from the incompressible $\mu(I)$-rheology to the compressible $\mu(J)$-$\phi(J)$-rheology to account for pore pressure, dilatancy, and the scaling laws under subaquatic conditions.
The model is supplemented with critical state theory to yield the correct properties in the quasi-static limit.
The pore fluid is described by an additional set of conservation equations and the interaction with grains is described by a drag model.
This new implementation enables us to include most of the physical processes relevant for submerged granular flows in a highly transparent manner.
Both, the incompressible and compressible rheologies are implemented into OpenFOAM and various simulations at low and high Stokes numbers are conducted with both frameworks.
We found a good agreement of the $\mu(J)$-$\phi(J)$-rheology with low Stokes number experiments, that incompressible models fail to describe.
The combination of granular rheology, pore pressure, and drag model leads to complex phenomena such as apparent cohesion, remoulding, hydroplaning, and turbidity currents.
The simulations give remarkable insights into these phenomena and increase our understanding of subaquatic mass transports.
\end{abstract}


\section{Introduction}
\label{sec:introduction}

Avalanches and landslides, as well as many industrial processes can be classified as granular flows.
Substantially improved rheological formulations have given rise to numerous attempts to simulate these phenomena with Navier-Stokes type models.
The vast amount of studies relies on the $\mu(I)$-rheology and its derivatives.
The core of the $\mu(I)$-rheology is the Drucker-Prager yield criterion \citep{drucker1952soil, rauter2020granular} and the recognition that the friction coefficient $\mu$ is solely a function of the inertial number $I$ \citep{midi2004dense, jop2006constitutive}.
Further studies found a similar correlation between the inertial number and the packing density $\phi$ \citep{forterre2008flows}.

A similar scaling was found in granular flows with low Stokes numbers $St$ (see Eq.~\eqref{eq:stokes}).
The Stokes number is related to the ratio between inertia and drag force on a particle and thus describes the influence of ambient fluid on the granular flow dynamics \citep[e.g.][]{finlay2001mechanics}.
Small Stokes numbers indicate a strong influence of the pore fluid on the particles, and hence also on the landslide dynamics.
In this regime, the viscous number $J$ replaces the inertial number $I$ as a control parameter for the friction coefficient $\mu$ and the packing density $\phi$, forming the so-called $\mu(J)$-$\phi(J)$-rheology \citep{boyer2011unifying}.
Furthermore, excess pore pressure can be remarkably high under these conditions and it is imperative to explicitly consider it in numerical simulations.
High drag forces and respectively small Stokes numbers are usually related to small particles.
They are virtually omnipresent in geophysical flows: submarine landslides \citep{kim2019landslide}, turbidity currents \citep{heerema2020determines}, powder snow avalanches \citep{sovilla2015structure}, and pyroclastic flows \citep{druitt1998pyroclastic} can be dominated by fine grained components.
It follows that a large portion of gravitational mass flows occurs at low Stokes numbers and a deeper understanding of the respective processes is relevant for many researchers.

Incompressible granular flow models have been applied in different forms to various problems in the last decade.
\cite{lagree2011granular} were the first to conduct numerical simulations of subaerial granular collapses with the $\mu(I)$-rheology and the finite volume method.
\cite{staron2012granular} used the same method to simulate silo outflows, and \cite{domnik2013coupling} used a constant friction coefficient to simulate granular flows on inclined plates.
\cite{vonboetticher2016debrisintermixing, vonboetticher2017debrisintermixing} applied a similar model, based on OpenFOAM, to debris flows and many more examples can be found in the literature.
More recently, compressible flow models have been introduced to simulate subaquatic granular flows at low Stokes numbers.
The applied methods include, e.g., smoothed particle hydrodynamics \citep{wang2017two}, coupled lattice Boltzmann and discrete element method \citep{yang2017role}, the material point method \citep{baumgarten2019general} or the finite volume multiphase framework of OpenFOAM \citep{si2018development}.
Results have often been compared to experiments of \cite{balmforth2005granular} (subaerial) and \cite{rondon2011granular} (subaquatic), two works that gained benchmark character in the granular flow community.

Most of the mentioned applications rely on standard methods from computational fluid dynamics (CFD).
This is reasonable, considering the similarity between the hydrodynamic (Navier-Stokes) equations and the granular flow equations.
However, the pressure dependent and shear thinning viscosity associated with granular flows introduces considerable conceptual and numerical problems.
The unconditional ill-posedness of an incompressible granular flow model with constant friction coefficient was described by \cite{schaeffer1987instability} and the partial ill-posedness of the $\mu(I)$-rheology by \cite{barker2015well}. 
By carefully tuning the respective relations, \cite{barker2017partial} were able to regularize the $\mu(I)$-rheology for all but very high inertia numbers.
\cite{barker2017well} described a well-posed compressible rheology, incorporating the $\mu(I)$-rheology as a special case.

Another pitfall of granular rheologies is the concept of effective pressure.
When pore pressure is considerably high (i.e. at low Stokes numbers), it is imperative to distinguish between effective pressure and total pressure \citep[first described by][]{terzaghi1925erdbaumechanik}.
Effective pressure represents normal forces in the grain skeleton that have a stabilizing effect, in contrast to pore pressure which has no stabilizing effect.
This has shown to be a major issue, as pore pressure and consequently the effective pressure, react very sensitively to the packing density and dilatancy \citep{rondon2011granular}.

Besides the rheology, tracking of the slide geometry poses a major challenge.
Surface tracking is usually implemented in terms of the algebraic volume-of-fluid (VOF) method \citep[e.g.][]{lagree2011granular, si2018development}, the level-set method \citep[e.g.][]{savage2014modeling}, geometric surface tracking methods \citep[e.g.][]{roenby2016computational, maric2018enhanced}, or particles based methods \citep[e.g.][]{baumgarten2019general, wang2017two}.


The volume-of-fluid method, which is also used in this work, allows to track the slide as a single component but also as a mixture of multiple phases (grains and pore fluid).
Components are defined in here as objects (e.g.~the landslide) that completely cover a bounded region in space without mixing with other components (e.g.~the ambient fluid), see Fig.~\ref{fig:alpha_def1}.
The tracking becomes a purely geometric problem \citep[see e.g.][for a geometric interpretation]{roenby2016computational}.
In contrast, phases (e.g.~grains) are dispersed and mixed with other phases (e.g.~pore fluid) to represent the dynamic bulk of the landslide, see Fig.~\ref{fig:alpha_def2}.

The component-wise tracking is used in various landslide models \citep[e.g.][]{lagree2011granular, domnik2013coupling, barker2017partial}.
Components, i.e.~the slide and the surrounding fluid, are immiscible and separated by a sharp interface.
Usually, this also implies that the model is incompressible.
The phase-wise tracking is commonly applied in chemical engineering \citep{gidaspow1994multiphase, vanwachem2000derivation, passalacqua2011implementation} and has lately been introduced to environmental engineering \citep[e.g.][]{cheng2017sedfoam, chauchat2017sedfoam, si2018development}.
This approach allows to describe a variable mixture of grains and pore fluid that merges smoothly into the ambient fluid.
The description of the pore fluid as an individual phase enables the model to decouple effective pressure from pore pressure, which is imperative in many flow configurations, e.g.~for low Stokes numbers.

In this work, a two-component and a two-phase Navier-Stokes type model are applied to granular flows.
Both models are implemented into the open-source toolkit OpenFOAM \citep{weller1998tensorial,rusche2002computational,opencfd2009user}, using the volume-of-fluid method for component- and phase-wise tracking (see section~\ref{sec:method}).
Subaerial \citep{balmforth2005granular} and subaquatic granular collapses \citep{rondon2011granular} are simulated with both models and results are compared to the respective experiments and with each other.

We apply the $\mu(I)$-$\phi(I)$-rheology to subaerial cases ($St \gtrapprox 1$) and the $\mu(J)$-$\phi(J)$-rheology to subaquatic cases ($St \lessapprox 1$).
The two-component model applies simplified rheologies in form of the incompressible $\mu(I)$- and $\mu(J)$-rheologies.
The $\phi(I)$- and $\phi(J)$-curves are merged into the particle pressure relation of \cite{johnson1987frictional} to achieve the correct quasi-static limits \citep{vescovi2013from}.
This yields reasonable values for the packing density at rest which is imperative for granular collapses with static regions.
In contrast to many previous works \citep[e.g.][]{savage2014modeling, vonboetticher2017debrisintermixing, si2018development}, we renounce additional contributions to shear strength (e.g.~cohesion) because we do not see any physical justification (e.g.~electrostatic forces, capillary forces, cementing) in the investigated cases.
We apply a very transparent and simple model, focusing on the relevant physical processes and achieve a remarkable accuracy, especially in comparison to more complex models \citep[e.g.][]{si2018development, baumgarten2019general}.
Further, it is shown that various experimental setups with different initial packing densities can be simulated with the same constitutive parameters, whereas many previous attempts required individual parameters for different cases \citep[e.g.][]{savage2014modeling, wang2017two, si2018development}.

The paper is organised as follows:
The multi-phase (section~\ref{ssec:twophase}) and multi-component (section~\ref{ssec:onephase}) models are introduced in section~\ref{sec:method}, including models for granular viscosity (section~\ref{ssec:rheo}), granular particle pressure (sections~\ref{ssec:ps1} and \ref{ssec:ps2}) and drag (section~\ref{ssec:drag}).
Results are shown and discussed in section~\ref{sec:balmforth} for a subaerial case and in section~\ref{sec:rondon} for two subaquatic cases.
A conclusion is drawn in section~\ref{sec:conclusion} and a summary is given in section~\ref{sec:summary}.
Furthermore, a thorough sensitivity analysis is provided in the appendix.

\section{Methods}
\label{sec:method}

\subsection{Two-phase landslide-model}
\label{ssec:twophase}

\begin{figure}
\begin{center}
\includegraphics[scale=1]{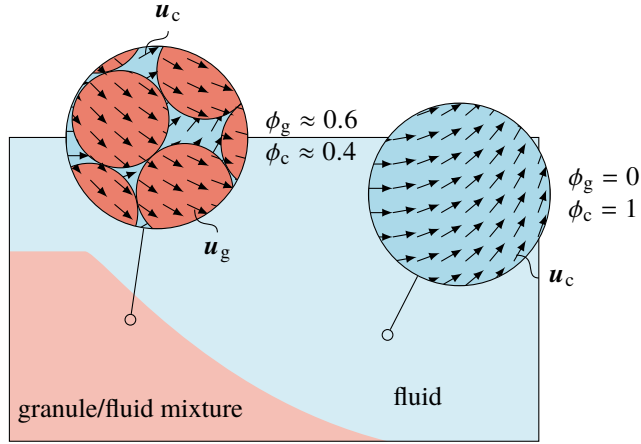}
\end{center}
\caption{Definition of phase-fractions $\phi_i$ and phase velocities $\b{u}_i$ in and outside a dense granular avalanche for the two-phase model. Phase velocities can differ, allowing phase-fractions to change, giving the avalanche compressible properties.}
\label{fig:alpha_def2}
\end{figure}

The two-phase model is based on the phase momentum and mass conservation equations \cite[see e.g.][]{rusche2002computational}.
The governing equations for the continuous fluid phase are given as
\begin{eqnarray}
&\dfrac{\partial \phi_{\r{c}}}{\partial t} + \bnabla\bcdot\left(\phi_{\r{c}}\,\b{u}_{\r{c}}\right) = 0,\label{eq:disp_alpha}\\
&\dfrac{\partial \phi_{\r{c}}\,\rho_{\r{c}}\,\b{u}_{\r{c}}}{\partial t} + \bnabla\bcdot\left(\phi_{\r{c}}\,\rho_{\r{c}}\,\b{u}_{\r{c}}\otimes\b{u}_{\r{c}}\right) = \bnabla\bcdot\left(\phi_{\r{c}}\,\bt{T}_{\r{c}}\right)-\phi_{\r{c}}\,\bnabla\,p+\phi_{\r{c}}\,\rho_{\r{c}}\,\b{g}+
k_{\r{gc}}\left(\b{u}_{\r{g}}-\b{u}_{\r{c}}\right).\label{eq:disp_momentum}
\end{eqnarray}
and for the grains as
\begin{eqnarray}
&\dfrac{\partial \phi_{\r{g}}}{\partial t} + \bnabla\bcdot\left(\phi_{\r{g}}\,\b{u}_{\r{g}}\right) = 0,\label{eq:disp_alpha_s}\\
&\dfrac{\partial \phi_{\r{g}}\,\rho_{\r{g}}\,\b{u}_{\r{g}}}{\partial t} + \bnabla\bcdot\left(\phi_{\r{g}}\,\rho_{\r{g}}\,\b{u}_{\r{g}}\otimes\b{u}_{\r{g}}\right) = \bnabla\bcdot\left(\phi_{\r{g}}\,\bt{T}_{\r{g}}\right)-\bnabla\,p_{\r{s}}-\phi_{\r{g}}\,\bnabla\,p+\phi_{\r{g}}\,\rho_{\r{g}}\,\b{g}+k_{\r{gc}}\left(\b{u}_{\r{c}}-\b{u}_{\r{g}}\right),\label{eq:disp_momentum_s}
\end{eqnarray}
Phase-fraction fields $\phi_{\r{g}}$ and $\phi_{\r{c}}$, i.e.~the phase volume over the total volume
\begin{equation}
\phi_{i} = \dfrac{V_{i}}{V},
\end{equation}
describe the composition of the grain-fluid mixture, see Fig.~\ref{fig:alpha_def2} (the index $i$ indicates either $\r{c}$ or $\r{g}$).
The granular phase-fraction is identical with the packing density $\phi = \phi_{\r{g}}$.
Phase-fractions take values between zero and one and the sum of all phase-fractions yields one.
The pore fluid is assumed to match the surrounding fluid and the respective phase-fraction $\phi_{\r{c}}$ is therefore one outside the slide.
This way, phase-fraction fields provide not only a mechanism to track the packing density of the slide, but also its geometry.
Every phase moves with a unique velocity field $\b{u}_{i}$, which is not divergence-free.
This allows the mixture to change, yielding a variable packing density and thus bulk-compressibility, although phase densities $\rho_{\r{g}}$ and $\rho_{\r{c}}$ are constant.
The volume weighted average velocity is divergence free,
\begin{equation}
\bnabla\bcdot \ol{\b{u}} = \bnabla\bcdot \left(\phi_{\r{g}}\,\b{u}_{\r{g}} + \phi_{\r{c}}\,\b{u}_{\r{c}} \right) = 0,\label{eq:divergencefreevel}
\end{equation}
which allows to use numerical methods for incompressible flow.

\begin{figure}
\begin{center}
\includegraphics[scale=1]{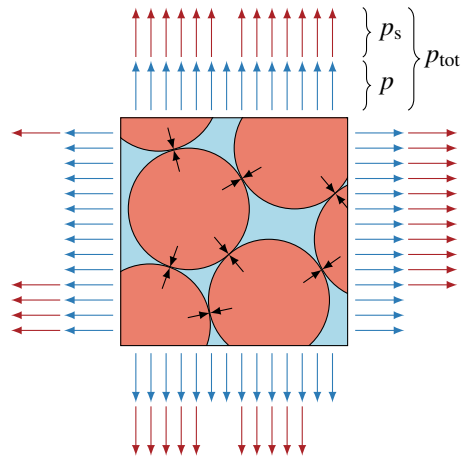}
\end{center}
\caption{Representative volume element of a grain-fluid mixture.
The effective pressure $p_{\r{s}}$ (red arrows) represents normal forces in the grain skeleton (black arrows). The pore-pressure (blue arrows) represents pressure that is equally shared by pore fluid and grains.}
\label{fig:particle_pressure}
\end{figure}

The pore pressure (or shared pressure) $p$ is acting on all phases equally, while the grain phase experiences additional pressure due to force chains between particles, the so called effective pressure (or particle pressure) $p_{\r{s}}$, see Fig.~\ref{fig:particle_pressure}.
The effective pressure is a function of the packing density in this model and the balance between effective pressure and external pressure (e.g.~overburden pressure) ensures realistic packing densities.
The total pressure can be assembled as
\begin{equation}
p_{\r{tot}} = p + p_{\r{s}}.\label{eq:p_tot}
\end{equation}
The deviatoric phase stress tensors are expressed as
\begin{equation}
\bt{T}_i = 2\,\rho_i\,\nu_i\bt{S}_i,\label{eq:def_viscosity_compressiblity}
\end{equation}
with phase viscosity $\nu_i$, phase density $\rho_i$ and deviatoric phase strain rate tensor 
\begin{equation}
\bt{S}_i = \dfrac{1}{2}\left(\bnabla\b{u}_i + \left(\bnabla\b{u}_i\right)^T\right) - \dfrac{1}{3}\bnabla \bcdot \b{u}_i\,\b{I.}
\end{equation}
The viscosity of the pore fluid $\nu_{\r{c}}$ is usually constant and the granular viscosity $\nu_{\r{g}}$ is following from constitutive models like the $\mu(I)$-rheology (see section~\ref{ssec:rheo}). 
The total deviatoric stress tensor can be calculated as
\begin{equation}
\bt{T} = \phi_{\r{c}}\,\bt{T}_{\r{c}} + \phi_{\r{g}}\,\bt{T}_{\r{g}}.\label{eq:total_dev}
\end{equation}

The last terms in Eqs.~\eqref{eq:disp_momentum} and \eqref{eq:disp_momentum_s} represent drag forces between phases and $k_{\r{gc}}$ is the drag coefficient of the grains in the pore fluid.
Lift and virtual mass forces are neglected in this work, because they play a minor role \citep{ si2018development}.

The granular viscosity $\nu_{\r{g}}$, the effective pressure $p_{\r{s}}$, and the drag coefficient $k_{\r{gc}}$ represent interfaces to exchangeable sub-models, presented in sections~\ref{ssec:rheo}-\ref{ssec:drag}.

\subsection{Two-component landslide-model}
\label{ssec:onephase}

Many two-phase systems can be substantially simplified by assuming that phases move together, i.e.~that phase velocities are equal,
\begin{equation}
\b{u}_{i} \approx \ol{\b{u}} = \phi_{\r{g}}\,\b{u}_{\r{g}} + \phi_{\r{c}}\,\b{u}_{\r{c}}.\label{eq:samevelocities}
\end{equation}
This fits very well to completely separated phases that are divided by a sharp interface \citep[e.g.~surface waves in water,][]{rauter2021numerical} but also systems of mixed phases (e.g.~grains and fluid) can be handled to some extent \citep[e.g.][]{lagree2011granular}.
The phase momentum conservation equations \eqref{eq:disp_momentum} and \eqref{eq:disp_momentum_s} can be combined into a single momentum conservation equation and the system takes the form of the ordinary Navier-Stokes Equations with variable fluid properties \citep[see e.g.][]{rusche2002computational},
\begin{eqnarray}
&\dfrac{\partial \rho\,\ol{\b{u}}}{\partial t} + \bnabla\bcdot\left(\rho\,\ol{\b{u}}\otimes\ol{\b{u}}\right) = \bnabla\bcdot\bt{T}-\bnabla\,p_{\r{tot}}+\rho\,\b{g},\label{eq:NS_momentum}\\
&\bnabla\bcdot\ol{\b{u}} = 0.\label{eq:NS_cont}
\end{eqnarray}
A detailed derivation can be found in appendix~\ref{sec:derivation}.
The pressure is denoted as $p_{\r{tot}}$, indicating that it contains contributions from hydrodynamic and effective pressure.

\begin{figure}
\begin{center}
\includegraphics[scale=1]{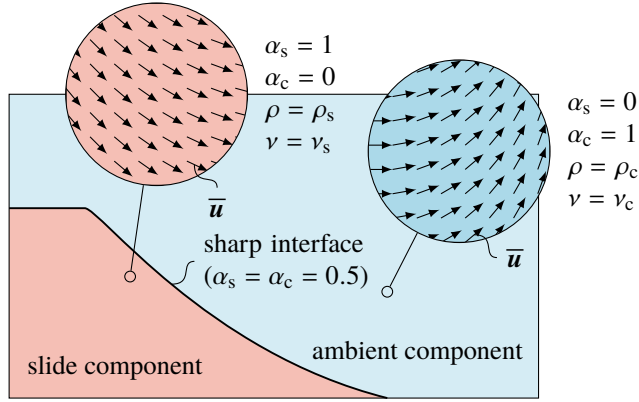}
\end{center}
\caption{Definition of component indicator functions $\alpha_i$ and the velocity $\b{\ol{u}}$ in and outside a dense granular avalanche for the two-component model.}
\label{fig:alpha_def1}
\end{figure}

The phase-fraction fields $\phi_i$ cannot be recovered after this simplification and the method switches to the tracking of components instead of phases, see Fig.~\ref{fig:alpha_def1}.
Components are tracked with so-called component indicator functions $\alpha_i$ (sometimes called phase indicator functions but in here we consequently distinguish phases from components), being either one if component $i$ is present at the respective location or zero otherwise,
\begin{equation}
\alpha_i = 
\begin{cases}
1 &\text{if component $i$ is present}\\
0 &\text{otherwise}
\end{cases}
\end{equation}
Values between zero and one are not intended by this method and only appear due to numerical reasons, i.e.~the discretisation of the discontinuous field (see section~\ref{ssec:numerics}).
In here, two component indicator functions are used, one for the ambient fluid component, $\alpha_{\r{c}}$, and one for the slide component, $\alpha_{\r{s}}$ (see Fig.~\ref{fig:alpha_def1}).
Evolution equations for component indicator functions can be derived from mass conservation equations as
\begin{equation}
\dfrac{\partial \alpha_i}{\partial t} + \bnabla\bcdot\left(\alpha_i\,\ol{\b{u}}\right) = 0.\label{eq:NS_alpha}
\end{equation}

The definition of components is straight forward for completely separated phases, where components can be matched with phases, e.g.~water and air.
The definition of the slide component, on the other hand, is not unambiguous, as it consists of a variable mixture of grains and pore fluid.
A boundary of the slide component can, for example, be found by defining a limit for the packing density (e.g.~50\% of the average packing density).
Further, a constant reference packing density $\ol{\phi}$ has to be determined, which is assigned to the whole slide component.
The density of the slide component follows as
\begin{equation}
\rho_{\r{s}} = \ol{\phi}\rho_{\r{g}} + (1-\ol{\phi})\rho_{\r{c}},\label{eq:mean_density}
\end{equation}
and a similar relation can be established for the deviatoric stress tensor (see section \ref{ssec:unifying}).

The local density $\rho$ and the local deviatoric stress tensor $\bt{T}$ can be calculated as
\begin{eqnarray}
&\rho = \sum\limits_i \alpha_i\,\rho_i = \alpha_{\r{s}}\,\rho_{\r{s}} + \alpha_{\r{c}}\,\rho_{\r{c}}\\
&\bt{T} = \sum\limits_i \alpha_i\,\bt{T}_i =\alpha_{\r{s}}\,\bt{T}_{\r{s}} + \alpha_{\r{c}}\,\bt{T}_{\r{c}},
\end{eqnarray}
using component densities $\rho_{i}$, as well as  component deviratoric stress tensors $\bt{T}_{i}$.
Component deviatoric stress tensors are calculated as
\begin{equation}
\bt{T}_{i} = \nu_{i}\,\rho_{i}\,\bt{S},
\end{equation}
with the component viscosity $\nu_i$ and the deviatoric shear rate tensor $\bt{S}$.
Note that the deviatoric shear rate tensor $\bt{S}$ matches the shear rate tensor $\bt{D}$, because the volume weighted averaged velocity field is divergence free,
\begin{equation}
\bt{S} = \bt{D} = \dfrac{1}{2}\left(\bnabla\,\b{\ol{u}}+\bnabla\,\b{\ol{u}}^T\right).
\end{equation}
The viscosity of the ambient fluid $\nu_{\r{c}}$ is usually constant and the viscosity of the slide region $\nu_{\r{s}}$ is following from granular rheology, see section~\ref{ssec:rheo}.

\subsection{Rheology}
\label{ssec:rheo}

\subsubsection{Unifying rheologies}
\label{ssec:unifying}

Most granular rheologies (e.g.~the $\mu(I)$-rheology) are defined in terms of the total deviatoric stress tensor in the slide component $\bt{T}_{\r{s}}$.
This has to be accounted for and corrected in the two-phase model if the same viscosity model is used in both models.
Similar to Eq.~\eqref{eq:mean_density}, component viscosities can be related to phase viscosities as
\begin{eqnarray}
\bt{T}_{\r{s}} = \ol{\phi}\bt{T}_{\r{g}} + (1-\ol{\phi})\,\bt{T}_{\r{c}},\\
2\,\rho_{\r{s}}\,\nu_{\r{s}}\bt{S}_{\r{s}} = 2\,\ol{\phi}\rho_{\r{g}}\,\nu_{\r{g}}\,\bt{S}_{\r{g}} + 2\,(1-\ol{\phi})\rho_{\r{c}}\,\nu_{\r{c}}\,\bt{S}_{\r{c}}. \label{eq:which_nu}
\end{eqnarray}
The contribution of the granular phase to stresses is assumed to be much higher than the contribution of the pore fluid, $\ol{\phi}\rho_{\r{g}}\,\nu_{\r{g}}\,\bt{S}_{\r{g}} \gg (1-\ol{\phi})\rho_{\r{c}}\,\nu_{\r{c}}\,\bt{S}_{\r{c}}$.
Further, by neglecting the mass of the pore fluid, $\rho_{\r{s}} \approx \ol{\phi}\,\rho_{\r{g}}$, it follows that kinematic viscosities have to be similar in both models,
\begin{equation}
\nu_{\r{s}} \approx \nu_{\r{g}}.
\end{equation}
Alternatively, one can match the dynamic viscosities $\nu_{\r{s}}\,\rho_{\r{s}}$ and $\nu_{\r{g}}\,\rho_{\r{g}}$ if the factor $\phi_{\r{g}}$ is removed from the viscous term in Eq.~\eqref{eq:disp_momentum_s}.
Note, that this assumptions are fairly accurate for subaerial granular flows but questionable for subaquatic granular flows.
However, multi-phase and multi-component models differ substantially under subaquatic conditions and a unification is not possible.

\subsubsection{Drucker-Prager plasticity model}
\label{sssec:coulomb}

An important characteristic of granular materials is the pressure dependent shear stress, described by the Drucker-Prager yield criterion \citep{drucker1952soil}.
\cite{schaeffer1987instability} was the first to include granular friction in the Navier-Stokes equations by expressing the Drucker-Prager yield criterion in terms of the shear rate tensor and the pressure,
\begin{equation}
\bt{T}_{\r{s}} = \mu\,p_{\r{s}}\,\dfrac{\bt{S}}{\|\bt{S}\|},\label{eq:granular}
\end{equation}
where the norm of a tensor $\|\bt{A}\|$ is defined as
\begin{equation}
\|\bt{A}\| = \sqrt{\dfrac{1}{2}\,\r{tr}\left(\bt{A}^2\right)}.\label{eq:thisnorm}
\end{equation}
The friction coefficient $\mu$ is constant and a material parameter in the first model by \cite{schaeffer1987instability}.
The slide component viscosity follows as 
\begin{equation}
\nu_{\r{s}} = \dfrac{\|\bt{T}_{\r{s}}\|}{\rho_{\r{s}}\,\|\bt{S}\|} = \mu\,\dfrac{p_{\r{s}}}{\rho_{\r{s}}\|\bt{S}\|}.\label{eq:granularviscosity}
\end{equation}
This relation has been applied with slight modifications by e.g.~\cite{domnik2013coupling}, \cite{savage2014modeling} or \cite{rauter2020granular}.
Following the findings in section~\ref{ssec:unifying}, the kinematic viscosity of slide and grains have to be similar and the granular phase viscosity follows as
\begin{equation}
\nu_{\r{g}} = \dfrac{\|\bt{T}_{\r{g}}\|}{\rho_{\r{s}}\,\|\bt{S}_{\r{g}} \|} = \mu\,\dfrac{p_{\r{s}}}{\rho_{\r{g}}\ol{\phi}\,\|\bt{S}_{\r{g}}\|}.\label{eq:granularviscosity2}
\end{equation}
The viscosity reaches very high values for $\|\bt{S}\| \rightarrow 0$ and very small values for $p_{\r{s}} \rightarrow 0$ and both limits can lead to numerical problems.
To overcome numerically unstable behaviour the viscosity is truncated to an interval $[\nu_{\min}, \nu_{\max}]$.
A thoughtful choice of $\nu_{\max}$ is crucial for the presented method.
Small values tend towards unphysical results, because solid-like behaviour can only be simulated by very high viscosities.
Big values, on the other hand, tend towards numerical instabilities (see section~\ref{sssec:timestepping}).
The ideal value for the maximum viscosity depends on the respective case and can be estimated with a scaling and sensitivity analysis (see appendix~\ref{ssec:convergence_viscosity}).
The relation
\begin{equation}
\nu_{\r{max}} = \dfrac{1}{10}\,\sqrt{|\b{g}|\,H^3},\label{eq:nu_thresh}
\end{equation}
where $H$ is the characteristic height of the investigated case, was found to give a good estimate for a reasonable viscosity cut-off.
Notably, the Drucker-Prager yield surface leads to an ill-posed model \citep{schaeffer1987instability} and the truncation of the viscosity is not sufficient for a regularization.

\cite{schaeffer1987instability} did not distinguish between effective and total pressure in Eq.~\eqref{eq:granularviscosity}, limiting the applications of his model substantially.
We will explicitly consider effective pressure in Eqs.~\eqref{eq:granularviscosity} and \eqref{eq:granularviscosity2} using Eq.~\eqref{eq:ps1} or \eqref{eq:ps2} in the two-component model and Eq.~\eqref{eq:ps3}, \eqref{eq:ps4}, or \eqref{eq:ps5} in the two-phase model to avoid such limitations.

\subsubsection{$\mu(I)$-rheology}
\label{sssec:muI}

The $\mu(I)$-rheology \citep{midi2004dense, jop2006constitutive, forterre2008flows} states that the friction coefficient $\mu$ is not constant in dense, dry, granular flows but rather a function of the inertial number $I$.
The inertial number $I$ is defined as the ratio between the typical time scale for microscopic rearrangements of grains with diameter $d$, $t_{\r{micro}} = d\,\sqrt{\rho_{\r{g}}/p_{\r{s}}}$, and the macroscopic time scale of the deformation, $t_{\r{macro}} =1/2\,\|\bt{S}\|^{-1}$,
\begin{equation}
I = 2\,d\,\|\bt{S}\| \sqrt{\dfrac{\rho_{\r{g}}}{p_{\r{s}}}},
\end{equation}
In the two-phase model, the shear rate $\bt{S}$ is replaced with the deviatoric shear rate of grains $\bt{S}_{\r{g}}$.
Various approaches have been proposed for the $\mu(I)$-curve, in here we apply the classic relation, given as
\begin{equation}
\mu(I) = \mu_{\r{1}} + \left(\mu_{\r{2}}-\mu_{\r{1}}\right)\dfrac{I}{I_0+I},\label{eq:muI}
\end{equation}
where $\mu_{\r{1}}$, $\mu_{\r{2}}$ and $I_0$ are material parameters \citep{jop2006constitutive}.
The dynamic friction coefficient $\mu(I)$ is introduced into the Drucker-Prager yield criterion, Eqs.~\eqref{eq:granularviscosity} or \eqref{eq:granularviscosity2} to get the respective granular viscosity.

\subsubsection{$\mu(J)$-rheology}
\label{sssec:muJ}

At small Stokes numbers, defined as
\begin{equation}
St = 2\,d^2\,\|\bt{S}\|\,\dfrac{\rho_{\r{g}}}{\nu_{\r{c}}\,\rho_{\r{c}}},\label{eq:stokes}
\end{equation}
the pore fluid has substantial influence on the rheology and the microscopic time scale is defined by the viscous scaling $t_{\r{micro}} = \nu_{\r{c}}\,\rho_{\r{c}}/p_{\r{s}}$ \citep{boyer2011unifying}.
The friction coefficient is thus no longer a function of the inertial number $I$ but rather of the viscous number $J$, defined as
\begin{equation}
J = 2\,\|\bt{S}\|\dfrac{\nu_{\r{c}}\,\rho_{\r{c}}}{p_{\r{s}}}.
\end{equation}
The functional relation of the friction coefficient on the viscous number was described by \cite{boyer2011unifying} as
\begin{equation}
\mu(J) = \mu_{\r{1}} + \left(\mu_{\r{2}}-\mu_{\r{1}}\right)\dfrac{J}{J_0+J} + J + \dfrac{5}{2}\,\phi_{\r{m}}\,\sqrt{J},\label{eq:muJ}
\end{equation}
where $\mu_{\r{1}}$, $\mu_{\r{2}}$, $J_0$ and $\phi_{\r{m}}$ are material parameters \citep{boyer2011unifying}.
The $\mu(J)$-rheology is taking advantage of the Drucker-Prager yield criterion, similar to the $\mu(I)$-rheology.

Notably, the $\mu(I)$ and $\mu(J)$-rheology can be combined by forming a new dimensionless number $K = J + \alpha\,I^2$ with a constitutive parameter $\alpha$ \citep{trulsson2012transition, baumgarten2019general}.
However, this was not required for the cases presented in this work.

\subsection{Effective pressure in the two-component model}
\label{ssec:ps1}

\subsubsection{Total pressure assumption}
\label{sssec:totalpressure}

The two-component model is limited in considering pore pressure and dilatancy effects because the packing density is not described by this model.
The effective pressure can only be reconstructed from total pressure $p_{\r{tot}}$ and various assumptions.
The simplest model assumes that the pore pressure is negligibly small, leading to
\begin{equation}
p_{\r{s}} \approx p_{\r{tot}}.\label{eq:ps1}
\end{equation}
This assumption is reasonable for subaerial granular flows and has been applied to such by e.g.~\cite{lagree2011granular} or \cite{savage2014modeling}.

\subsubsection{Hydrostatic pressure assumption}
\label{sssec:hydrostatic}

In subaquatic granular flows, the surrounding high-density fluid increases the total pressure substantially and it cannot be neglected.
Following \cite{savage2014modeling}, improvement can be achieved by calculating the hydrostatic pore pressure as
\begin{equation}
p_{\r{hs}} = \begin{cases} \rho_{\r{c}}\,\b{g}\bcdot\left(\b{x} - \b{x}_0\right) & \text{for} \quad \b{g}\bcdot\left(\b{x} - \b{x}_0\right) > 0,\\
0 & \text{else},
\end{cases}
\end{equation}
and subtracting it from the total pressure,
\begin{equation}
p_{\r{s}} \approx p_{\r{tot}} - p_{\r{hs}}.\label{eq:ps2}
\end{equation}
Here, $\b{x}_0$ is the position of the free water surface, where the total pressure is supposed to be zero.
For a variable and non-horizontal free water surface, common in e.g.~landslide-tsunamis, this concept is complicated substantially, and to the authors knowledge, not applied.
Furthermore, excess pore pressure, which is common in low Stokes number flows, is out of the scope for this model.

\subsection{Effective pressure in the two-phase model}
\label{ssec:ps2}

\subsubsection{Critical state theory}
\label{sssec:cst}

The structure of the two-phase model allows us to include the packing density in the effective pressure equation.
Critical state theory \citep{roscoe1958yielding, roscoe1970influence, schofield1968critical} was the first model to describe the relationship between the effective pressure and the packing density.
The critical state is defined as a state of constant packing density and constant shear stress, which is reached after a certain amount of shearing of an initially dense or loose sample.
The packing density in this state, called critical packing density $\phi_{\r{crit}}$, is a function of the effective pressure $p_{\r{s}}$.
This function can be inverted to get the effective pressure as a function of the critical packing density.
It is further assumed that the flow is in its critical state $\phi_{\r{g}} = \phi_{\r{crit}}$ to get a model that is compatible with the governing equations.
This assumption is reasonable for avalanches, slides, and other granular flows but questionable for the initial release and deposition.
At small deformations, the packing density might be lower (underconsolidated) or higher (overconsolidated) than the critical packing density and the effective pressure model will over- or underestimate the effective pressure.

A popular relation for the effective pressure (the so-called critical state line) has been described by \cite{johnson1987frictional, johnson1990frictional} as
\begin{equation}
p_{\r{s}} = a\,\dfrac{\phi_{\r{g}}-\phi_{\r{rlp}}}{\phi_{\r{rcp}}-\phi_{\r{g}}},\label{eq:ps3}
\end{equation}
where $\phi_{\r{rlp}}$ is the random loose packing density in critical state, $\phi_{\r{rcp}}$ the random close packing density in critical state and $a$ a scaling parameter.
The scaling parameter $a$ can be interpreted as the effective pressure at the packing density $\frac{1}{2}\left(\phi_{\r{rcp}}+\phi_{\r{rlp}}\right)$.
Note that we apply a simplified version of the original relation, similar to \cite{vescovi2013from}.
Packing densities above $\phi_{\r{rcp}}$ are not valid and avoided by the asymptote of the effective pressure at $\phi_{\r{rcp}}$.
If packing densities higher or equal $\phi_{\r{rcp}}$ appear in simulations, they should be terminated and restarted with refined numerical parameters (e.g.~time step duration).

\subsubsection{$\phi(I)$-relation}
\label{sssec:phiI}

\begin{figure*}
\begin{center}
\includegraphics[scale=0.75]{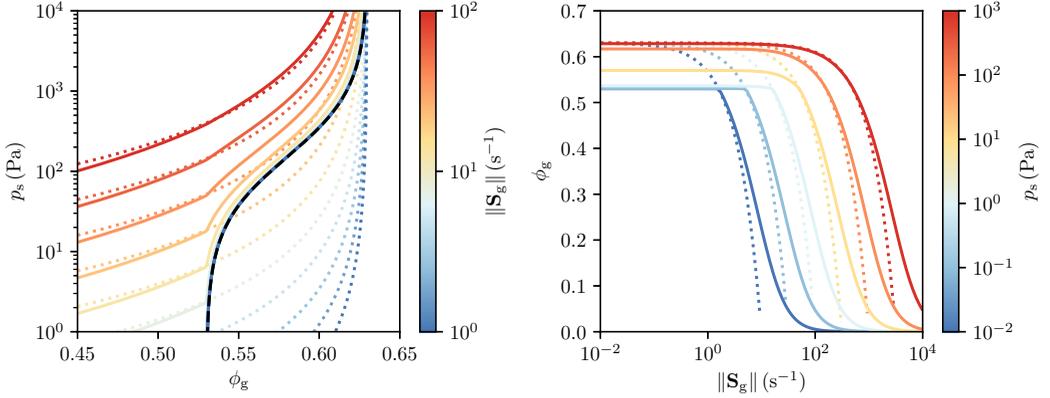}
\end{center}
\caption{Left: Effective pressure $p_{\r{s}}$ following the $\phi(I)$-relation as a function of packing density $\phi_{\r{g}}$ and deviatoric shear rate $\|\bt{S}_{\r{g}}\|$. The dashed lines show the original relation of \cite{forterre2008flows}, the continuous coloured lines show the modified relation and the black line the quasi-static limit following \cite{johnson1987frictional}. Right: The critical packing density as a function of particle pressure $p_{\r{s}}$ and deviatoric shear rate $\|\bt{S}_{\r{g}}\|$. Dashed lines are following the original $\phi(I)$-relation, continuous lines the modified version. The critical state theory would result in horizontal lines in this plot.}
\label{fig:alphai}
\end{figure*}

Equation~\eqref{eq:ps3} is known to hold for slow deformations in critical state \citep[see e.g.][]{vescovi2013from}.
However, this relation is not consistent with granular flow experiments.
Granular flows show dilatancy with increasing shear rate, expressed by e.g.~\cite{forterre2008flows} as a function of the inertia number $I$,
\begin{equation}
\phi_{\r{g}}(I) = \phi_{\max} -\Delta\phi\,I,
\end{equation}
where $\phi_{\max}$ and $\Delta\phi$ are  material parameters.
This relation can be transformed into a model for the effective pressure by introducing the inertial number $I$,
\begin{equation}
p_{\r{s}} = \rho_{\r{s}}\,\left(2\,\|\bt{S}_{\r{g}}\|\,d\,\dfrac{\Delta\phi}{\phi_{\max}-\phi_{\r{g}}}\right)^{2}.\label{eq:ps4.0}
\end{equation}
This relation has two substantial problems: For $\|\bt{S}_{\r{g}}\| = 0$ it yields $p_{\r{s}} = 0$ and for $\phi_{\r{g}} = 0$ it yields $p_{\r{s}} \neq 0$, which causes numerical problems and unrealistic results.
The first problem is addressed by superposing Eq.~\eqref{eq:ps4.0} with the quasi-static relation \eqref{eq:ps3}, similar to \cite{vescovi2013from}.
The second problem is solved by multiplying Eq.~\eqref{eq:ps4.0} with the normalized packing density $\phi_{\r{g}}/\ol{\phi}$, which ensures that the pressure vanishes for $\phi_{\r{g}} = 0$.
The normalization with the reference packing density $\ol{\phi}$ ensures that parameters ($\phi_{\max}$, $\Delta\phi$) will be similar to the original equation.
Further, to reduce the number of material parameters, we set the maximum packing density in the $\phi(I)$-relation equal to the random close packing density $\phi_{\r{rcp}}$.
The final relation reads
\begin{equation}
p_{\r{s}} = a\,\dfrac{\phi_{\r{g}}-\phi_{\r{rlp}}}{\phi_{\r{rcp}}-\phi_{\r{g}}} 
+ \rho_{\r{g}}\,\dfrac{\phi_{\r{g}}}{\ol{\phi}}\left(2\,\|\bt{S}_{\r{g}}\|\,d\,\dfrac{\Delta\phi}{\phi_{\r{rcp}}-\phi_{\r{g}}}\right)^{2},\label{eq:ps4}
\end{equation}
and is shown in Fig.~\ref{fig:alphai} alongside the original relations of \cite{johnson1987frictional} and \cite{forterre2008flows}.
Interestingly, this relation contains many features of the extended kinetic theory of \cite{vescovi2013from} (compare Fig.~\ref{fig:alphai}b with Fig. 6b in \cite{vescovi2013from}).
Notably, the inertial number is a function of only the packing density and the shear rate, $I = f\left(\phi_{\r{g}}, \|\bt{S}_{\r{g}}\|\right)$, because the effective pressure is calculated as function of the packing density. The same follows for the friction coefficient $\mu = f\left(\phi_{\r{g}}, \|\bt{S}_{\r{g}}\|\right)$ and the deviatoric stress tensor $\|\bt{T}_{\r{g}}\| = f\left(\phi_{\r{g}}, \|\bt{S}_{\r{g}}\|\right)$.
This highlights that the two-phase model implements a density-dependent rheology, rather than a pressure-dependent rheology.

It should be noted that there are various possibilities to combine critical state theory and the $\mu(I)$-$\phi(I)$-rheology.
An alternative approach including bulk viscosity is provided by e.g.~\cite{schaeffer2019constitutive}.

\subsubsection{$\phi(J)$-relation}
\label{sssec:phiJ}

\begin{figure*}
\begin{centering}
\includegraphics[scale=0.75]{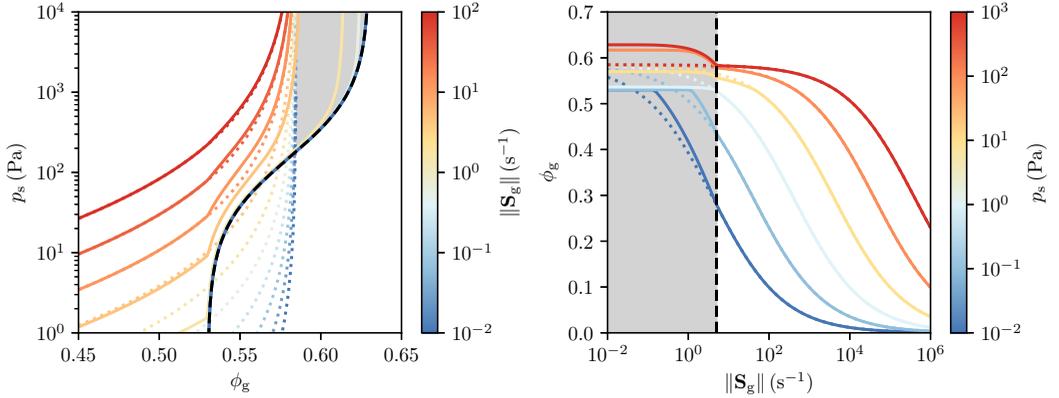}
\end{centering}
\caption{Left: Particle pressure $p_{\r{s}}$ following the $\phi(J)$-relation as a function of packing density $\phi_{\r{g}}$ and deviatoric shear rate $\|\bt{S}_{\r{g}}\|$. The dashed lines show the original relation of \cite{boyer2011unifying}, the continuous coloured lines show the modified relation and the black line the static limit expressed following \cite{johnson1987frictional}. Right: The critical packing density as a function of particle pressure $p_{\r{s}}$ and deviatoric shear rate $\|\bt{S}_{\r{g}}\|$. Dashed lines are following the original $\phi(J)$-relation, continuous lines the modified version. The grey area shows the region where only creeping shear rates below $S_0$ are allowed.}
\label{fig:alphaj}
\end{figure*}

The low Stokes number regime requires the replacement of the inertial number $I$ with the viscous number $J$.
The dependence of the packing density on the viscous number was described by \cite{boyer2011unifying} as
\begin{equation}
\phi_{\r{g}} = \dfrac{\phi_{\r{m}}}{1+\sqrt{J}},
\end{equation}
and we can derive the effective pressure by inserting the viscous number as
\begin{equation}
p_{\r{s}} = \dfrac{2\,\nu_{\r{c}}\,\rho_{\r{c}}\,\|\bt{S}_{\r{g}}\|}{\left(\frac{\phi_{\r{m}}}{\phi_{\r{g}}}-1\right)^2}.
\end{equation}
Notably, \cite{boyer2011unifying} emphasised that $\phi_{\r{m}}$ is not matching the random close packing density $\phi_{\r{rcp}} \approx 0.63$ but rather a value close to $0.585$.
This leads to substantial problems for large values of $\phi_{\r{g}}$ as the relation is only valid for $\phi_{\r{g}} < \phi_{\r{m}} = 0.585 $ or $\|\bt{S}_{\r{g}}\| = 0$.
In other words, shearing is only possible for $\phi_{\r{g}} < \phi_{\r{m}}$.
We solve this issue by allowing a creeping shear rate of $S_0$ at packing densities above $\phi_{\r{m}}$.
Further and as before, we superpose the relation with the quasi-static relation of \cite{johnson1987frictional} to yield the correct asymptotic values for $\|\bt{S}_{\r{g}}\| \rightarrow 0$ known from critical state theory.
The final relation reads
\begin{equation}
p_{\r{s}} = a\,\dfrac{\phi_{\r{g}}-\phi_{\r{rlp}}}{\phi_{\r{rcp}}-\phi_{\r{g}}} + \dfrac{2\,\nu_{\r{c}}\,\rho_{\r{c}}\,\|\bt{S}_{\r{g}}\|}{\left(\frac{\hat{\phi}_{\r{m}}}{\phi_{\r{g}}}-1\right)^2},\label{eq:ps5}
\end{equation}
with
\begin{equation}
\hat{\phi}_{\r{m}} = 
\begin{cases} 
\phi_{\r{m}}+\left(\phi_{\r{rcp}}-\phi_{\r{m}}\right)\,\left(S_{0}-\|\bt{S}\|\right) & \text{for} \quad S_{0} > \|\bt{S}\|, \\
\phi_{\r{m}} & \text{else}.
\end{cases}
\end{equation}
The respective relation is shown in Fig.~\ref{fig:alphaj} alongside the original relations of \cite{johnson1987frictional} and \cite{boyer2011unifying}.
States with $\|\bt{S}\| \geq S_0$ and $\phi_{\r{g}} \geq \phi_{\r{m}}$ or $\phi_{\r{g}} \geq \phi_{\r{rcp}}$ are not intended by this model and simulations should be terminated if such states appear.

\subsection{Drag and permeability model}
\label{ssec:drag}

\begin{figure*}
\begin{centering}
\includegraphics[scale=0.75]{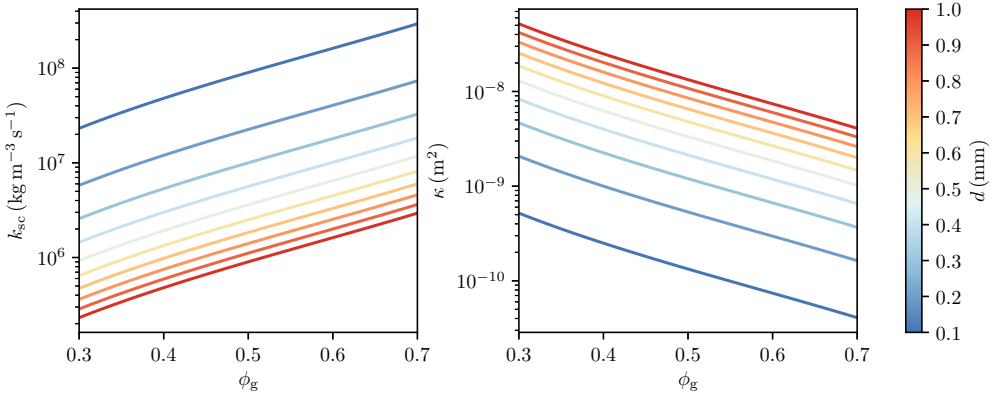}
\end{centering}
\caption{Drag coefficient $k_{\r{gc}}$ (left) and permeability $\kappa$ (right) following the Kozeny-Carman relation \citep{pailha2009two} for various grain diameters (colour) and packing densities (x-axis).}
\label{fig:drag}
\end{figure*}

The drag model describes the momentum exchange between grains and pore fluid in the two-phase model and widely controls permeability, excess pore pressure relaxation, and the settling velocity of grains.
A wide range of drag models for various situations can be found in the literature.
In here we stick to the Kozeny-Carman relation as applied by \cite{pailha2009two},
\begin{equation}
k_{\r{g}\r{c}} = 150\,\dfrac{\phi_{\r{g}}^2\,\nu_{\r{c}}\,\rho_{\r{c}}}{\phi_{\r{c}}\,d^2},\label{eq:ksc}
\end{equation}
with the grain diameter $d$ as the only parameter.
This relation is supposed to be valid for small relative velocities and densely packed granular material.
It has been modified to account for higher relative velocities \citep{ergun1952fluid} and lower packing densities \citep{gidaspow1994multiphase}, however, which is not relevant for the investigated configurations (see \cite{si2018development} for an application of the extended relation).
This relation is visualized in Fig.~\ref{fig:drag}a for various diameters and packing densities.

The drag coefficient can be reformulated into a permeability coefficient as known in soil mechanics and porous media.
Comparing Darcy's law \citep[e.g.][]{bear1972dynamics} with the equations of motion for the fluid phase, we can calculate the hydraulic conductivity as 
\begin{equation}
K = \dfrac{\rho_{\r{c}}|\b{g}|}{k_{\r{gc}}}
\end{equation}
and furthermore the intrinsic permeability \citep[e.g.][]{bear1972dynamics} as
\begin{equation}
\kappa = K\dfrac{\nu_{\r{c}}}{|\b{g}|} = \dfrac{\nu_{\r{c}}\,\rho_{\r{c}}}{k_{\r{gc}}} = \dfrac{\phi_{\r{c}}\,d^2}{150\,\phi_{\r{g}}^2}.\label{eq:permeability}
\end{equation}
The permeability is visualized in Fig.~\ref{fig:drag}b.
In a similar manner, the drag coefficient can be calculated as 
\begin{equation}
k_{\r{g}\r{c}} = \dfrac{\rho_{\r{c}}\,\nu_{\r{c}}}{\kappa},
\end{equation}
if the intrinsic permeability of the granular material is known.

\subsection{Numerical solution and exception handling}
\label{ssec:numerics}

All models are implemented into OpenFOAM-v1812 \citep{weller1998tensorial, opencfd2009user} and solved with the finite volume method \citep{jasak1996error, rusche2002computational, moukalled2016finite}.

\subsubsection{Two-component landslide-model}

The two-component model is based on the solver multiphaseInterFoam, using the PISO-algorithm \citep{issa1986solution} and interpolations following \cite{rhie1983numerical} to solve the coupled system of pressure and velocity.
First, an updated velocity field is calculated without the contribution of pressure.
The predicted velocity field is later corrected to be divergence-free and the pressure follows from the required correction.
Finally, all other fields, e.g.~the phase indicator functions, are updated.
This procedure is repeated in each time step.

Components (slide and ambient air or water) are divided by an interface which is supposed to be sharp.
However, the interface is often smeared by numerical diffusion.
To keep the interface between components sharp, the relative velocity between phases $\b{u}_{\r{ij}}$, which was previously eliminated from the system, is reintroduced in Eq.~\eqref{eq:NS_alpha},
\begin{equation}
\dfrac{\partial \alpha_i}{\partial t} + \bnabla\bcdot\left(\alpha_i\,\b{\ol{u}}\right) + \bnabla\bcdot\left(\alpha_i\,\alpha_j\,\b{u}_{ij}\right) = 0.\label{eq:sp_alpha}
\end{equation}
Eq.~\eqref{eq:sp_alpha} is finally solved using the MULES algorithm (Multidimensional Universal Limiter with Explicit Solution) \citep{weller2008new}.
This scheme limits the interface compression term (i.e.~the term containing $\b{u}_{ij}$) to avoid over- ($\alpha_i > 1$) and undershoots ($\alpha_i < 0$) of the component indicator fields.

There is no conservation equation for the relative velocity in the two-component model and it has to be reconstructed from assumptions.
Two methods are known to construct the relative velocity for granular flows.
\cite{barker2020coupling} suggest to construct the relative velocity for granular flows from physical effects such as segregation and settling.
The relative velocity follows as the terminal velocity of spheres in the surrounding fluid under the influence of gravity.
Alternatively, one can construct a velocity field that is normal to the interface and of the same magnitude as the average velocity $\b{\ol{u}}$,
\begin{equation}
\b{u}_{ij} = |\b{\ol{u}}|\dfrac{\alpha_{j}\,\bnabla\,\alpha_{i} - \alpha_{i}\,\bnabla\,\alpha_{j}}{|\alpha_{j}\,\bnabla\,\alpha_{i} - \alpha_{i}\,\bnabla\,\alpha_{j}|}.
\end{equation}
This method has a maximum sharpening effect \citep{weller2008new} and is thus also applied in this work.

\subsubsection{Two-phase landslide-model}

The two-phase model is based on the solver multiphaseEulerFoam.
The system of pressure and average velocity is solved with the same concept as in the two-component solver. 
The velocity fields for all phases are first predicted without contributions from pore pressure $p$, but including effective pressure $p_{\r{s}}$.
The average velocity is then corrected to be divergence-free and the pore pressure follows from the required correction.
In a further step, the velocity correction is applied to phase velocities.
The solution procedure is described in depth by \cite{rusche2002computational}.
The interface compression term is not required in this model because settling and segregation is directly simulated and counteracting numerical diffusion.
The implementation of the effective pressure term is taken from SedFoam~2.0 \citep{chauchat2017sedfoam}.


\subsubsection{Time stepping}
\label{sssec:timestepping}

The numerical solution of transport equations is subject to limitations that pose restrictions on the solution method.
One of these limitations is known as the Courant-Friedrichs-Lewy (CFL) condition and enforced by limiting the CFL number.
In convection dominated problems, the CFL number is defined as the ratio of the time step duration $\Delta t$ and the cell convection time $\Delta x/u_x$, i.e.~the time required for a particle to pass a cell with size $\Delta x$,
\begin{equation}
\r{CFL}^{\r{conv}} = \dfrac{u_x\,\Delta t}{\Delta x}.\label{eq:cfl_conv}
\end{equation}
For the stability of e.g.~the forward Euler method, it is required, that the convection time is smaller than the time step duration,
\begin{equation}
\r{CFL}^{\r{conv}} \leq 1,
\end{equation}
and similar limits exist for other explicit methods.
This limitation has to be enforced by choosing the time step duration $\Delta t$ according to mesh size and flow velocity.

However, Eq.~\eqref{eq:cfl_conv} is only valid for convection dominated problems.
In the case of granular flows, the viscosity term is dominating over all other terms.
Therefore, the viscosity has to be considered in the calculation of the CFL number and the time step duration.
The respective definition, ignoring the contribution of convection follows as
\begin{equation}
\r{CFL}^{\r{diff}} = \dfrac{\nu\,\Delta t}{\Delta x^2}.\label{eq:cfl_nu}
\end{equation}
This relation is imperative for stability of explicit and semi-implicit Navier-Stokes solvers when viscous forces are dominating.
The squared cell size in the denominator and the high viscosity introduce very strict limitations on the time step, making computations very expensive.
Note that simplified relations for the one-dimensional case are given in here.
The full multi-dimensional conditions for arbitrary finite volume cells can be found in \cite{rauter2021numerical}.

\section{Subaerial granular collapse \citep{balmforth2005granular}}
\label{sec:balmforth}

\begin{figure}
\begin{center}
\includegraphics[scale=1]{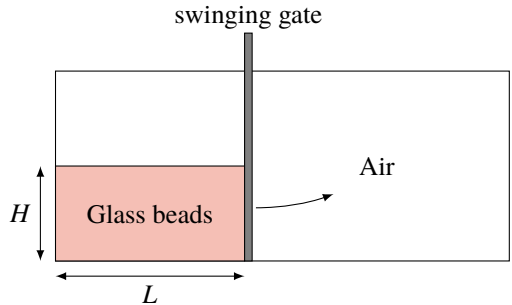}
\end{center}
\caption{Experimental column collapse setup of \cite{balmforth2005granular}. The aspect ration $H/L$ has been varied throughout the experiments. We will focus on the experiment $L=0.2\,\r{m}$, $H=0.1\,\r{m}$, similar to \cite{savage2014modeling}.}
\label{fig:balmforth}
\end{figure}

As a first test of the numerical models, we simulate the granular collapse experiments of \cite{balmforth2005granular} under subaerial conditions.
A sketch of the experiment is shown in Fig.~\ref{fig:balmforth}.
The experiment was conducted between two parallel, smooth walls and the setup is approximated as a 2D granular collapse.
\cite{balmforth2005granular} conducted multiple experiments with different geometries, in here we focus on the experiments with an aspect ratio of $H/L = 1/2$, but similar results have been obtained for other aspect ratios.
In theory, both, the two-component and the two-phase model should be equally capable of simulating this case because pore pressure plays a minor role.
Most parameters, such as density, quasi-static friction coefficient, and particle diameter are reported by \cite{balmforth2005granular}.
The missing parameters are completed with data from the literature.
Notably, the experiments are conducted on a smooth surface, which was incorporated in simulations by switching to a constant friction coefficient $\mu_{\r{wall}}$ at smooth surfaces.
This modification is simple in the finite volume method because stresses are calculated on cell faces before their divergence is calculated as a sum over faces.

The Stokes number is estimated to be of order $10^3$ (with $\|\bt{S}\| = 10\,\r{s^{-1}}$) for this experiments and the $\mu(I)$-$\phi(I)$-rheology is chosen to describe friction and effective pressure.
Parameters for the $\mu(I)$ and $\phi(I)$-curves are chosen in the physically reasonable range ($\mu_2-\mu_1 \approx 0.3$, $I_0 \approx 0.25$, $\Delta\phi=0.1$) following various references \citep[e.g.][]{forterre2008flows} in combination with values reported by \cite{balmforth2005granular}.
A wide range of limiting packing densities can be found in literature, $\phi_{\r{rlp}}$ varying between $0.5$ \citep{si2018development} and $0.598$ \citep{vescovi2013from}, $\phi_{\r{rcp}}$ varying between $0.619$ \citep{vescovi2013from} and $0.64$ \citep{savage2014modeling}.
These parameters are therefore optimized to the subaquatic case (section~\ref{sec:rondon}), where extended measurements are available, and applied to this case without further modification.
The average packing density is assumed to be $\ol{\phi} = 0.6$ following the critical state line at this pressure level.
The applied pressure equation is visualized in Fig.~\ref{fig:alphai}.
From the height $H=0.1\,\r{m}$ the required viscosity threshold $\nu_{\max}$ can be estimated following Eq.~\eqref{eq:nu_thresh} to be of order $1\,\r{m^{2}\,s^{-1}}$.
This estimation was validated with a sensitivity analysis (see appendix~\ref{ssec:convergence_viscosity}).
The final set of parameters is given in Tab.~\ref{tab:balmforth}.

\begin{table}
\caption{Material parameters for the subaerial granular collapse simulations. Note that not all material parameters are required by all models.}
\label{tab:balmforth}
\begin{center}
\begin{tabular}{llll}
\hline
phase / component    & par.                              & value                              & description\\
\hline
air      & $\rho_{\r{c}}$                    & $1\,\r{kg\,m^{-3}}$                & air density\\
         & $\nu_{\r{c}}$                     & $1.48\bcdot10^{-5}\,\r{m^{2}\,s^{-1}}$ & air viscosity\\
\hline
slide / grains    & $d$                               & $10^{-3}\,\r{m}$                   & particle diameter\\
         & $\mu_{\r{wall}}$                  & $0.317$                            & wall friction coefficient\\
         & $\mu_{\r{1}}$                     & $0.595$                            & quasi-static friction coefficient\\
         & $\mu_{\r{2}}$                     & $0.895$                            & dynamic friction coefficient\\
         & $I_0$                             & $0.25$                             & reference inertial number\\
         & $\nu_{\r{min}}$                   & $10^{-4}\,\r{m^{2}\,s^{-1}}$       & lower viscosity threshold\\
         & $\nu_{\r{max}}$                   & $1\,\r{m^{2}\,s^{-1}}$             & upper viscosity threshold\\
         & $\ol{\phi}$                       & $0.60$                             & assumed mean packing density$^3$\\
         & $\rho_{\r{s}}$                    & $1\,430\,\r{kg\,m^{-3}}$           & slide density$^1$\\
         & $\rho_{\r{g}}$                    & $2\,600\,\r{kg\,m^{-3}}$           & particle density$^2$\\
         & $\phi_{\r{rlp}}$                  & $0.53$                             & random loose packing density$^2$\\
         & $\phi_{\r{rcp}}$                  & $0.63$                             & random close packing density$^2$\\
         & $a$                               & $130\,\r{Pa}$                      & critical state line parameter$^2$\\
         & $\Delta\phi$                      & $0.1$                              & dynamic loosening factor$^2$\\
\hline
\end{tabular}
\end{center}
$^1$only two-component model.\\
$^2$only two-phase model.\\
$^3$used to match kinematic viscosity in the two-phase model following Eq.~\eqref{eq:which_nu}.\\
\end{table}

Regular meshes of square cells are used to cover a simulation domain of $0.5\times0.2\,\r{m}$, which was sufficient to have no artificial influences from boundaries.
Standard boundary conditions are applied at walls (zero velocity, zero pressure gradient) and the permeable top (zero velocity gradient, zero pressure).
Multiple mesh resolutions were applied to investigate the influence of the grid resolution on the results (see appendix~\ref{ssec:convergence_grid}).
The time stepping was investigated with a similar approach, modifying the limit for $\r{CFL}_{\r{max}}^{\r{diff}}$ between $1$ and $1000$ (depending on model and solver mode, see appendix~\ref{ssec:convergence_timestep}).
In the following, the CFL-number is limited to $1$ and the cell size set to $0.0017\,\r{m}$, which showed to be sufficient to achieve converged and mesh independent results.

\subsection{Two-component model}
\label{ssec:balmforth_tc}

The component indicator for the slide component $\alpha_{\r{s}}$ is initialized to $1$ within the square that forms the initial granular column.
We assume that hydrostatic pore pressure is negligible ($ < 2\,\r{Pa}$) and therefore apply Eq.~\eqref{eq:ps1} to calculate the effective pressure.

\begin{figure}
\begin{center}
\includegraphics[scale=1]{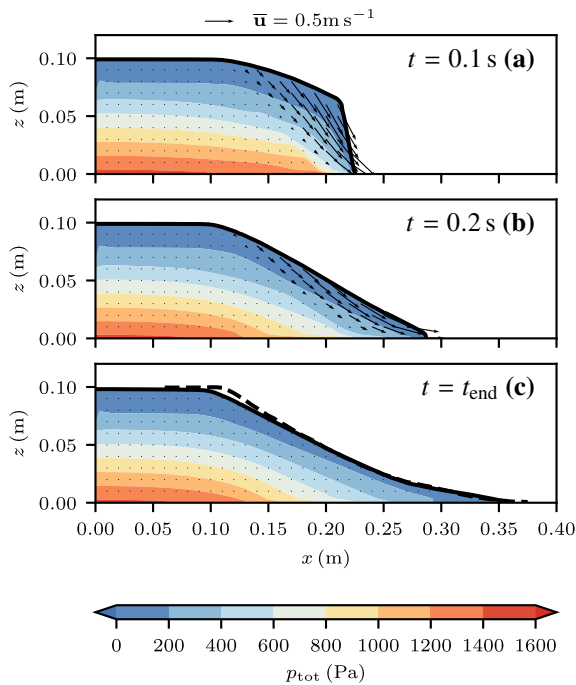}
\end{center}
\caption{
Total pressure, assumed to match the effective pressure in the two-component model (subaerial case).
The black arrows represent the velocity.
The continuous black line shows the free surface of the slide ($\alpha_{\r{s}} = 0.5$), the dashed black line shows the final experimental pile shape of \cite{balmforth2005granular}.
}
\label{fig:2005_sp_p}
\end{figure}

The simulation covering a simulation duration of $0.8\,\r{s}$ took $6.9\,\r{h}$ on eight cores of LEO4 (High Performance Cluster from the University of Innsbruck, consisting of Intel Xeon (Broadwell) compute cores).
The total pressure, which is assumed to match the effective pressure, is shown for three-time steps in Fig.~\ref{fig:2005_sp_p}, alongside the final pile in the experiment.
The continuous black line shows the sharp free surface, assumed to be located at $\alpha_{\r{s}} = 0.5$.
Furthermore, the velocity field $\b{\ol{u}}$ is shown as arrows.
The collapse takes about $0.4\,\r{s}$ and the pile remains in its final shape for the rest of the simulation.
The two-component model matches the experiment well, however, the volume of the final pile is slightly underestimated.
Results are very robust in terms of mesh refinement or coarsening (see appendix~\ref{ssec:convergence_grid}) and mesh dependent instabilities \cite[as e.g.][]{martin2017continuum, gesenhues2019finite} have not been observed.

\subsection{Two-phase model}
\label{ssec:balmforth_tp}

The two-phase model uses the same parameters as the two-component model, including numerical parameters, such as viscosity threshold and CFL limit.
The phase-fraction $\phi_{\r{g}}$ was initialized such that effective pressure is in balance with lithostatic vertical stresses, yielding an initial mean phase-fraction of $\overline{\phi_{\r{g}}} = 0.608$.
This procedure ensures that there will be no dilatancy or compaction in stable regions of the pile.

\begin{figure}
\begin{center}
\includegraphics[scale=1]{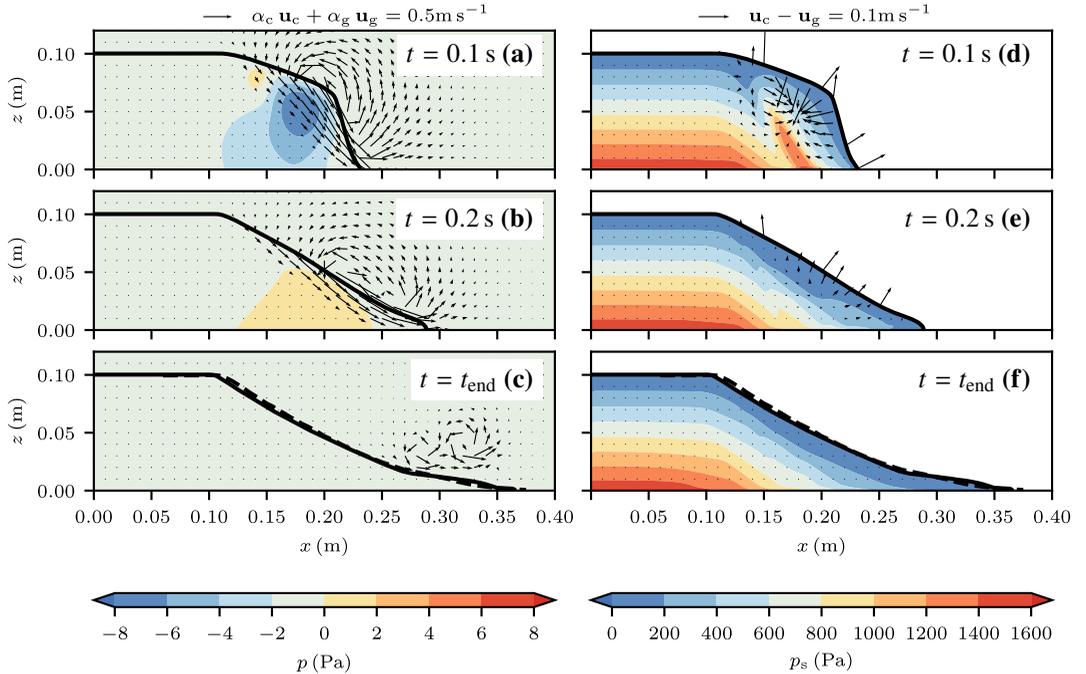}
\end{center}
\caption{
Pore pressure (a-c) and effective pressure (d-f) in the two-phase model (subaerial case). The arrows show the average velocity (a-c) and the relative velocity (d-f).
The continuous black line shows the free surface of the slide ($\phi_{\r{g}} = 0.25$), the dashed black line shows the final experimental pile shape of \cite{balmforth2005granular}.
}
\label{fig:2005_tp_p}
\end{figure}

The simulation took $9.1\,\r{h}$ under the same conditions as the two-component simulation.
A stronger mesh dependency is observed for this model, however, the runout is converging for fine meshes (see appendix~\ref{ssec:convergence_grid}).
The pore pressure and the effective pressure following the extended $\phi(I)$-theory are shown for three time steps in Fig.~\ref{fig:2005_tp_p}, alongside the final pile shape in the experiment.
The continuous black line indicates the position of the free surface, assumed to be located at $\phi_{\r{s}} = 0.25$.
The average velocity is shown as arrows in Figs.~\ref{fig:2005_tp_p}a-c, the relative velocity of air with respect to grains in Figs.~\ref{fig:2005_tp_p}d-f.
The relative velocity in the initial phase is considerably high, indicating an inflow of air into the bulk and thus dilatancy.
The two-phase model matches the experiment exceptionally well and the dilatancy in the experiment is matched by the simulation to a high degree.
Note that the effective pressure at rest is directly linked to the packing density which can be qualitatively estimated from Fig.~\ref{fig:2005_tp_p}f.

\subsection{Discussion and comparison}
\label{ssec:balmforth_dc}


Both models performed well at simulating the subaerial granular collapse.
This is in line with previous results of e.g.~\cite{lagree2011granular} or \cite{savage2014modeling}.
The effective pressure and the total pressure are fairly similar, because excess pore pressure is dissipating quickly through dilatancy or compaction.
The magnitude of pore pressure in the two-phase model is smaller than $8\,\r{Pa}$ and thus less than $1\,\%$ of the effective pressure, validating the assumption of neglectable pore pressure.


The runout is similar in both models, the front is slightly elongated in the two-phase model.
Further, the two-phase model shows a better match with the experiment at the upper end of the final slope.
Both of these minor differences can be attributed to dilatancy effects.
The two-component model is intrinsically not able to capture this process.
Two mechanisms for dilatancy can be observed in the two-phase model.
Firstly, the average effective pressure in the slide is reduced as it is spreading out and the packing density decreases proportionally to the effective pressure, as prescribed by the critical state line.
Secondly, shearing can reduce the packing density well below its critical packing density due to the dynamic contribution of the $\phi(I)$-theory to effective pressure.
The loosely packed slide will not return to the critical packing density after shearing but remain in a loose state, forming a hysteresis.
The granular material is able to remain in a loose state because the deviatoric stress tensor counteracts one-dimensional settling deformations (known as oedometric compression in soil mechanics).
Furthermore, the granular column may have been overconsolidated in the experiment, however, this was not incorporated in the model due to the initialisation in critical state.

Dilatancy is rather unimportant under subaerial conditions, as it does not imply changes in rheology or flow dynamics.
Therefore, the two-component model is well suited for subaerial granular collapses, where pore pressure is negligibly small and the Stokes number is well above one.


The reduced friction at the smooth basal surface has a small but noticeable effect on the final pile shape.
The runout is longer when incorporating the smooth surface and matches the experiment better.
Previous works \citep[e.g.][]{savage2014modeling} ignored the smooth bottom of the experiment and still obtained accurate final pile shapes by using a constant friction coefficient.
The increased friction of the $\mu(I)$-rheology (in comparison to a constant quasi-static friction coefficient) compensates for the reduced basal friction quite exactly (see appendix~\ref{ssec:dynamics}).



The two-component model is less sensitive to grid resolution than the two-phase model (see appendix \ref{ssec:convergence_grid}) but more sensitive to the time step duration (see appendix \ref{ssec:convergence_timestep}).
At the same resolution, both models require roughly the same computational resources and no model shows a substantial advantage in this regard.

It is important to carefully choose the time step duration, as it can have drastic influences on simulation results.
Generally, $\r{CFL}^{\r{diff}}$ has to be limited to one to guarantee satisfying results, while some cases and solver settings allow higher $\r{CFL}^{\r{diff}}$ numbers.
This limitation is much stronger than the traditional CFL criterion and $\r{CFL}^{\r{conv}}$ is roughly $0.001$.
Notably, the time step duration is constant in simulations, $\Delta t \approx 3\bcdot10^{-6}\,\r{s}$, because the constant maximum viscosity $\nu_{\r{max}}$ in stable regions and the constant cell size $\Delta x$ controlled the time stepping.

\section{Subaqueous granular collapse \citep{rondon2011granular}}
\label{sec:rondon}

\begin{figure}
\begin{center}
\includegraphics[scale=1]{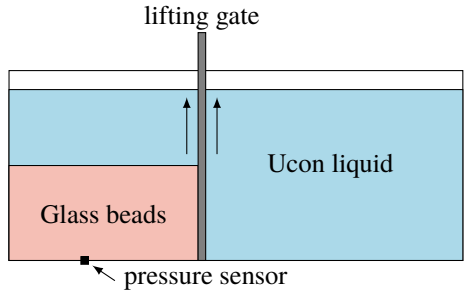}
\end{center}
\caption{Experimental column collapse setup of \cite{rondon2011granular}. The packing density and the aspect ratio have been varied in the experiment. We will focus on a densely and a loosely packed case, similar to \cite{savage2014modeling}.}
\label{fig:rondon}
\end{figure}

The granular collapse experiments of \cite{rondon2011granular} are conducted under subaquatic conditions and the Stokes number was estimated to be of order $10^{-1}$ (at $\|\bt{S}\| = 10\,\r{s^{-1}}$).
Pore pressure, packing density, and permeability play an important role under these conditions and the complexity increases substantially.
Experiments accounted for the increased complexity by varying the average initial packing density in experiments between $0.55$ and $0.61$.
The pore pressure was recorded by a sensor in the bottom plate, approximately below the centre of the column at $x=0.02\,\r{m}$ (see Fig.~\ref{fig:rondon}).
This sensor showed strong variations of the pore pressure in dense and loose experiments, indicating its important role for subaquatic slides.

A loose or underconsolidated ($\ol{\phi_{\r{g}}} = 0.55$, $L=0.06\,\r{m}$, $H=0.048\,\r{m}$) and a dense or overconsolidated ($\ol{\phi_{\r{g}}} =0.6$, $L=0.06\,\r{m}$, $H=0.042\,\r{m}$) simulation are conducted in this work to investigate the sensitivity of the model.
As before, the experiments were conducted between two parallel, smooth walls and the setup is approximated with 2D simulations.
Most material parameters are reported by \cite{rondon2011granular}, parameters for the $\mu(J)$ and $\phi(J)$-curves are supplemented with data from \cite{boyer2011unifying}.
The quasi-static friction coefficient $\mu_{\r{1}}$ is taken from \cite{si2018development}.
The particles have a diameter of  $d=0.225\,\r{mm}$ and are immersed into a Ucon solution \citep[for details, see][]{rondon2011granular} with a viscosity of $\nu_{\r{c}}=1.2\bcdot10^{-5}\,\r{m^{2}\,s^{-1}}$ (about $10$ times higher than water), leading to a very low permeability of $\kappa\approx10^{-10}\r{m^2}$ following Eq.~\eqref{eq:permeability}.
Early tests revealed that the two-phase model reacts very sensitively to the critical state line parameters $\phi_{\r{rlp}}$, $\phi_{\r{rcp}}$, and $a$.
Parameters from literature \citep[e.g.~the critical state line applied by][]{si2018development} lead to unrealistic granular pressures at $\phi_{\r{g}}=0.60$ and could thus not be applied.
We set the limiting packing densities to $\phi_{\r{rlp}}=0.53$ and $\phi_{\r{rcp}}=0.63$ to allow initial average packing densities between $0.55$ and $0.61$.
The scaling variable $a$ was found by matching the peak pore pressure in the dense simulation with the respective measurement (see Fig.~\ref{fig:2011time_p}).
The total set of parameters used for both cases is shown in Tab.~\ref{tab:rondon}.

Regular meshes of square cells with size $0.0005\,\r{m}$ are applied, covering a simulation domain of $0.15\,\r{m}\times0.105\,\r{m}$ (dense case) and $0.25\,\r{m}\times0.105\,\r{m}$ (loose case).
The CFL number $\r{CFL}^{\r{diff}}$ is limited to $10$ in order to keep computation times to a reasonable level.
A sensitivity study was conducted to proof convergence at this grid size (see appendix~\ref{ssec:convergence_grid}) and $\r{CFL}^{\r{diff}}$ number (see appendix~\ref{ssec:convergence_timestep}).

\begin{table}
\caption{Material parameters for the subaquatic granular collapse simulations. Note that not all material parameters are required by all models.}
\label{tab:rondon}
\begin{center}
\begin{tabular}{llll}
\hline
phase    & par.                          & value                              & description\\
\hline
ucon mix & $\rho_{\r{c}}$                    & $1\,000\,\r{kg\,m^{-3}}$           & ucon mix density\\
         & $\nu_{\r{c}}$                     & $1.2\bcdot10^{-5}\,\r{m^{2}\,s^{-1}}$   & ucon mix viscosity\\
\hline
slide    & $d$                               & $2.25\bcdot10^{-4}\,\r{m}$          & particle diameter\\
         & $\mu_{\r{1}}$                     & $0.340$                            & quasi-static friction coefficient\\
         & $\mu_{\r{2}}$                     & $0.740$                            & dynamic friction coefficient\\
         & $J_0$                             & $0.005$                            & reference viscous number\\
         & $\nu_{\r{min}}$                   & $10^{-4}\,\r{m^{2}\,s^{-1}}$           & lower viscosity threshold\\
         & $\nu_{\r{max}}$                   & $1\,\r{m^{2}\,s^{-1}}$                 & upper viscosity threshold\\
         & $\ol{\phi}$                       & $0.60$                             & assumed mean packing density$^3$\\
         & $\rho_{\r{s}}$                    & $1\,900\,\r{kg\,m^{-3}}$           & slide density$^1$\\
         & $\rho_{\r{g}}$                    & $2\,500\,\r{kg\,m^{-3}}$           & particle density$^2$\\
         & $\phi_{\r{rlp}}$                  & $0.53$                             & random loose packing density$^2$\\
         & $\phi_{\r{rcp}}$                  & $0.63$                             & random close packing density$^2$\\
         & $a$                               & $130\,\r{Pa}$                      & critical state line parameter$^2$\\
         & $\phi_{\r{m}}$                    & $0.585$                            & dynamic reference packing density$^2$\\
         & $S_0$                             & $5\,\r{s^{-1}}$                    & maximum creep shearing$^2$\\
\hline
\end{tabular}
\end{center}
$^1$only two-component model.\\
$^2$only two-phase model.\\
$^3$used to match kinematic viscosity in the two-phase model following Eq.~\eqref{eq:which_nu}.
\end{table}

\subsection{Two-component model - dense case}
\label{ssec:rondon_tc1}

\begin{figure}
\begin{center}
\includegraphics[scale=1]{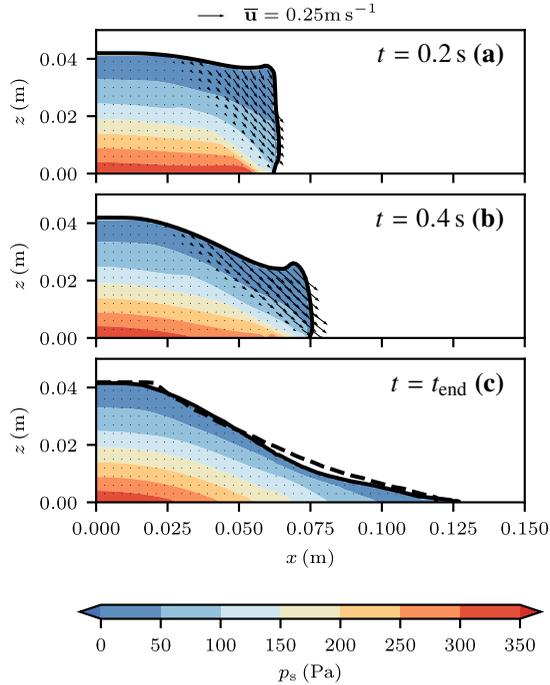}
\end{center}
\caption{
Effective pressure at $t=0.2\,\r{s}$ (a), $t=0.4\,\r{s}$ (b) and $t=1.0\,\r{s}$ (c) in the two-component model (subaquatic dense case). The black arrows represent the velocity.
The continuous black line shows the free surface of the slide ($\alpha_{\r{s}} = 0.5$), the dashed black line shows the final experimental pile shape of \cite{rondon2011granular}.
}
\label{fig:2011_sp_p1}
\end{figure}

The hydrostatic pore pressure is high under subaquatic conditions and the two-component model applies Eq.~\eqref{eq:ps2} to consider its influence on the effective pressure.
All parameters are taken from Tab.~\ref{tab:rondon}.
The evolution of the slide geometry, the effective pressure and the velocity $\b{\ol{u}}$ are shown in Fig.~\ref{fig:2011_sp_p1}, alongside the final experimental pile shape.
The final pile shape of the model corresponds roughly to the experiment.
The velocity, on the other hand, is roughly corresponding to the loose case, and the collapse is completed after $1\,\r{s}$, whereas the dense experiment took more than $30\,\r{s}$.
The simulation and its failure mechanism are similar to the subaerial case where the free unsupported side of the pile is collapsing until reaching a stable slope inclination.
Notably, neither the dense nor the loose experiment showed such a failure mechanism (see Fig.~\ref{fig:sketch}).
No excess pore pressure is included in this model and a hypothetical pressure sensor at the bottom of the column would measure constantly $0\,\r{Pa}$ as indicated in Fig.~\ref{fig:2011time_p}.

\subsection{Two-component model - loose case}
\label{ssec:rondon_tc2}

The two-component model provides only a few and ineffective possibilities to consider variations of the packing density.
To simulate the loose granular collapse with this model, the average packing density is changed to $\ol{\phi} = 0.55$ and the bulk density correspondingly to $\rho_{\r{s}} = 1825\,\r{kg\,m^3}$.
Further, the initial column geometry is changed as reported by \cite{rondon2011granular}.
All other parameters match the dense case.
Changing rheology parameters, e.g.~$\mu_1$ or $\mu_2$ \citep[as e.g.][]{wang2017two} is technically possible but does not help in understanding the physical process or the influence of packing density.

The difference to the dense simulation is very small and thus not shown in here \cite[see, e.g.][for similar results]{bouchut2017two}.
As before, the final pile shape is close to the dense experiment while the simulated velocity is close to the loose experiment.
The runout is slightly longer as in the dense simulation because the loose column is slightly taller.

\subsection{Two-phase model - dense case}
\label{ssec:rondon_tp1}

\begin{figure}
\includegraphics[scale=1]{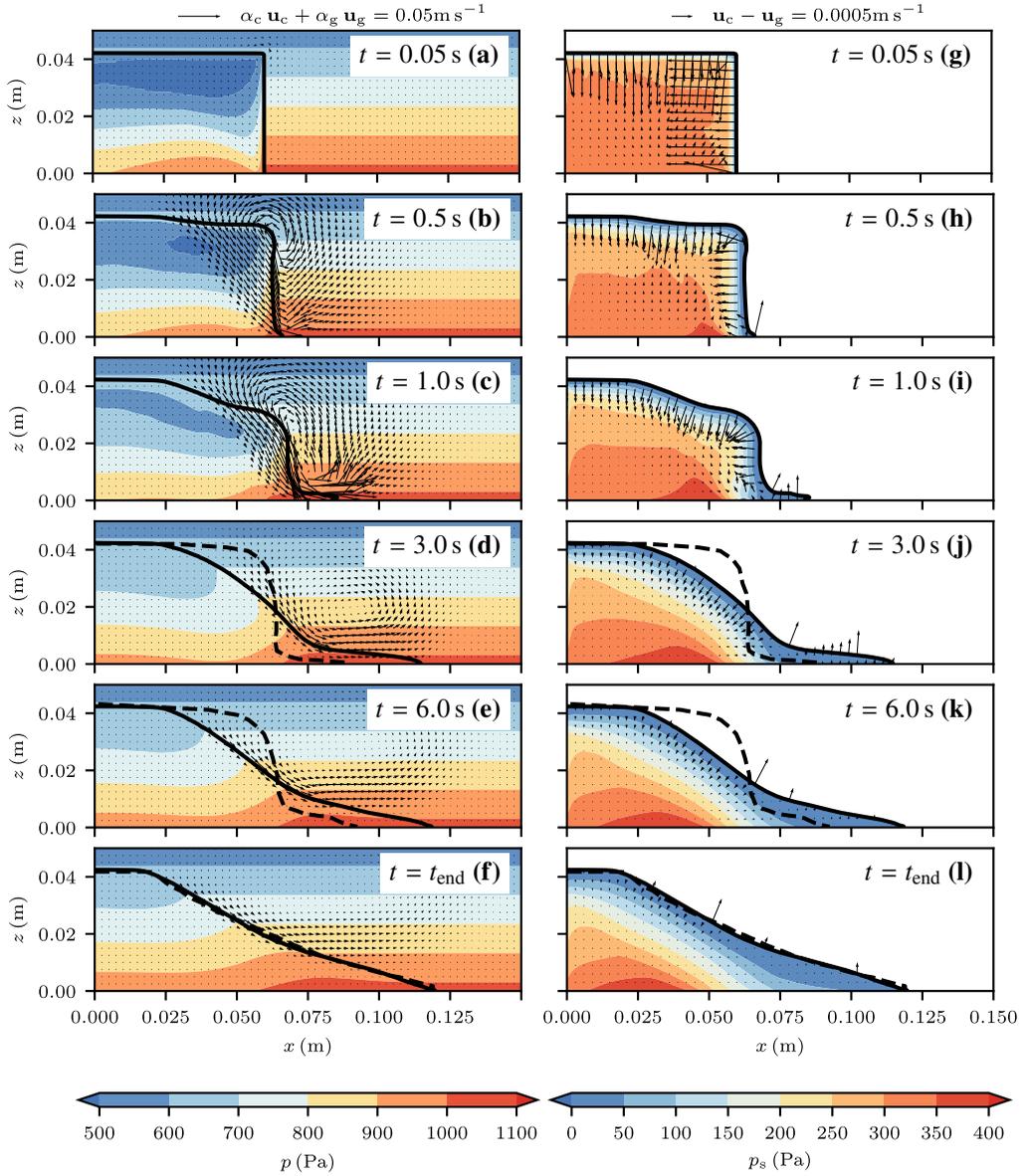}
\caption{
Pore pressure (a-f) and effective pressure (g-l) at $t=0.05\,\r{s}$ (a, g), $t=0.5\,\r{s}$ (b, h), $t=1\,\r{s}$ (c, i), $t=3\,\r{s}$ (d, j), $t=6\,\r{s}$ (e, k) and the final state (f,l) using the two-phase model (subaquatic dense case). The black arrows represent the average velocity (a-f) and the relative velocity (g-l). The final state ($t_{\r{end}}$) is reached at $t=10\,\r{s}$ in the simulation (small velocities remain) but $t=30\,\r{s}$ in the experiment. The black line shows the free surface of the slide, assumed at $\phi_{\r{s}}=0.25$. The free surface of the experiment is shown for comparison as a black dashed line.
}
\label{fig:2011d_p}
\end{figure}

The two-phase model allows us to explicitly consider variations in the initial packing density.
The dense case is initialized with a homogeneous packing density of $0.60$. 
The evolution of the dense granular column as simulated with the two-phase model is shown in Fig.~\ref{fig:2011d_p}, alongside three states of the experiment at $t=3\,\r{s}$, $6\,\r{s}$ and $30\,\r{s}$.
The simulation covering a duration of $10\,\r{s}$, took $240\,\r{h}$ on 8 cores of LEO4.
The dense case is dominated by negative excess pore pressure (Fig.~\ref{fig:2011d_p}a-e), meaning that pore pressure within the slide is lower than outside.
The effective pressure (Fig.~\ref{fig:2011d_p}f-j) is respectively higher, which increases the shear strength of the column.
Initially, the shear strength is high enough to delay the collapse and to keep the column mostly stable.
The pore pressure gradient leads to the suction of fluid into the column (Fig.~\ref{fig:2011d_p}g-h) and the granular material is dilating.
Dilation reduces the effective pressure and allows the column to collapse.
This happens first near the free surface on the unsupported side of the column, leading to a breaching-like flow of grains (Fig.~\ref{fig:2011d_p}g-h).
Grains mix with fluid at the breaching edge, reducing packing density, effective pressure, and thus friction to very low values.
The resulting mixture behaves like a small turbidity current and reaches long run-outs with shallow slopes, as visible in Fig.~\ref{fig:2011d_p}i-j.
The zone of low particle pressure extends towards the centre of the column with time and further mobilisation occurs.
At $t = 0.5\,\r{s}$, we can see the formation of a shear band.
The grains above the shear band slide off, first as a triangular cohesive block (note the uniform velocity field in Fig.~\ref{fig:2011d_p}b), which disintegrates between $t=1\,\r{s}$ and $t=3\,\r{s}$ (Fig.~\ref{fig:2011d_p}i). 
The overall process is finished (i.e.~$t_{\r{end}}$) in the simulation after roughly $10\,\r{s}$, while the experiment took about $30\,\r{s}$.
The final pile form and the failure mechanism match the experiment very well, which can be seen best in a comparison with the videos provided by \cite{rondon2011granular}, see Fig.~\ref{fig:sketch}.
Further, a good match with the measured excess pore pressure is achieved, as shown in Fig.~\ref{fig:2011time_p}.
The time scale and velocity of the collapse, on the other hand, differ substantially between simulation and experiment.
Notably, pore pressure $p$ and effective pressure $p_{\r{s}}$ do not sum up to the total vertical load, as a considerable fraction of the vertical load is transferred to the ground by viscous stresses.

\subsection{Two-phase model - loose case}
\label{ssec:rondon_tp2}

\begin{figure}
\includegraphics[scale=1]{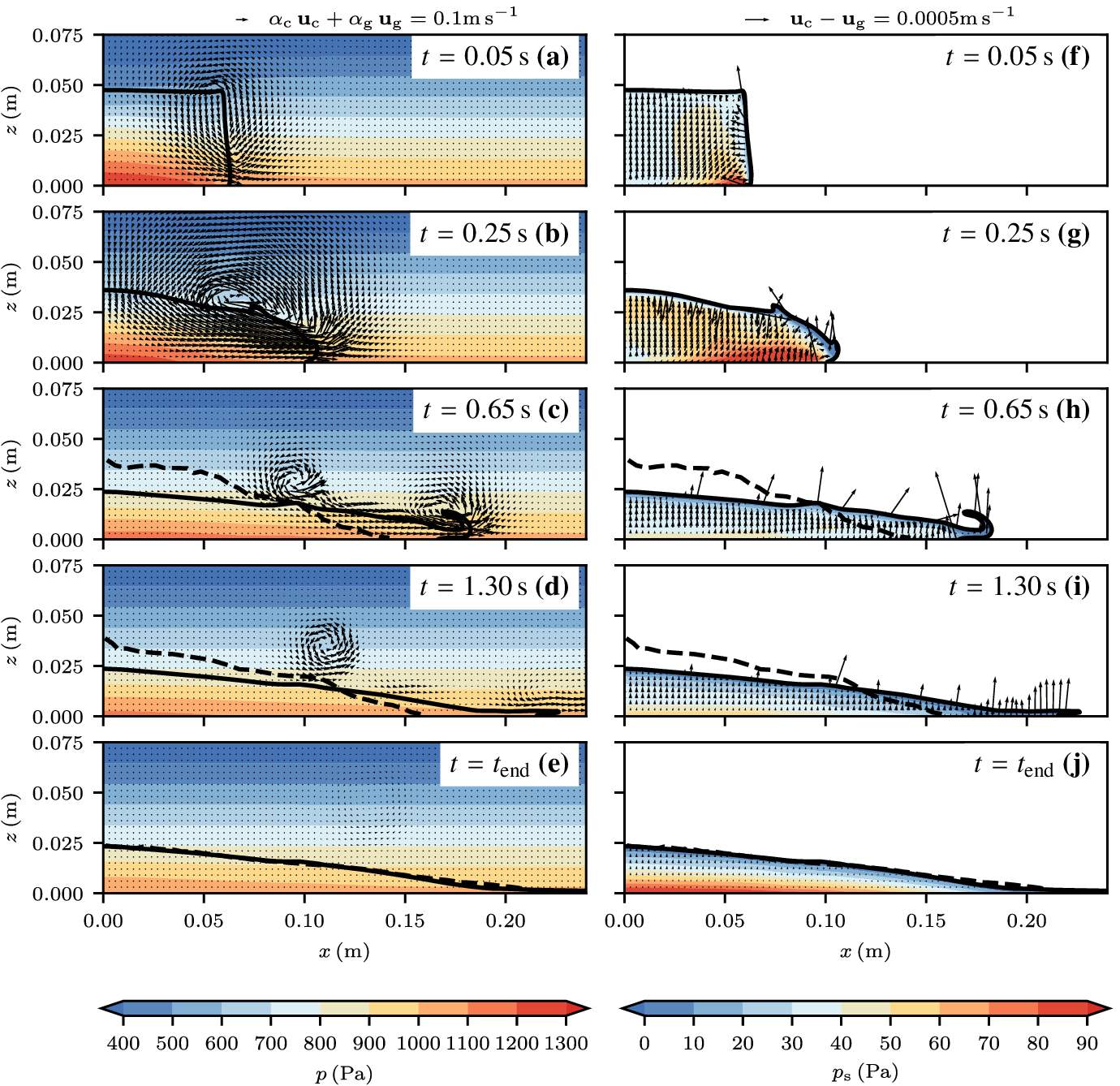}
\caption{Pore pressure (a-e) and effective pressure (f-j) at $t=0.05\,\r{s}$ (a, f), $t=0.25\,\r{s}$ (b, g) , $t=0.65\,\r{s}$ (c, h), $t=1.30\,\r{s}$ (d, i) and  the final state ($t_{\r{end}} = 6.0\,\r{s}$) (e, j) using the two-phase model (subaquatic loose case). 
The black arrows represent the average velocity (a-e) and the relative velocity (f-j). The black line shows the free surface of the slide, assumed at $\phi_{\r{s}}=0.25$. The free surface of the experiment is shown for comparison as a black dashed line.
}
\label{fig:2011l_p}
\end{figure}

\begin{figure}
\begin{center}
\includegraphics[scale=1]{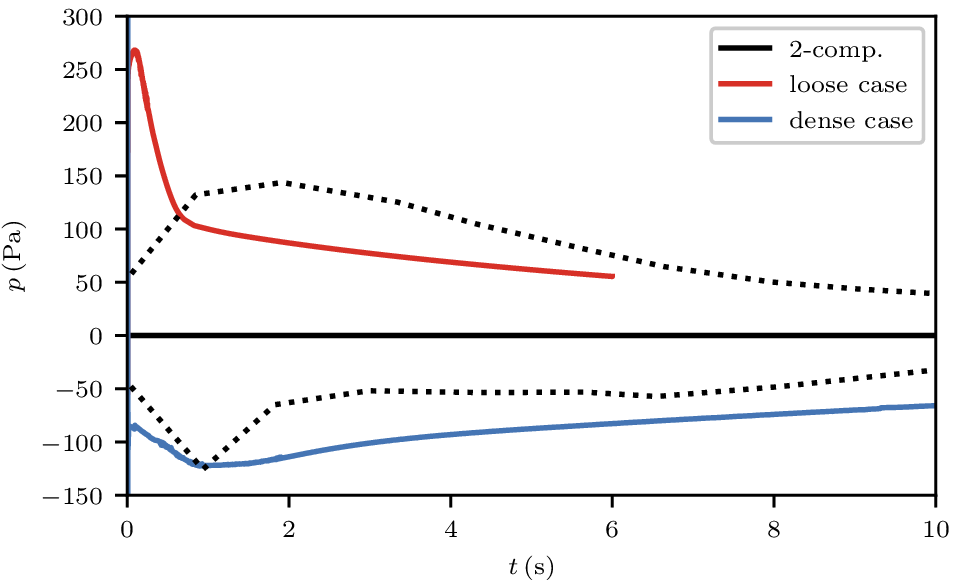}
\end{center}
\caption{The excess pore pressure as a function of time for the subaquatic granular collapses. 
The loose simulation (red) shows a strong peak of excess pore pressure that exceeds the experimental measurement (upper black dashed line).
The dense simulation (blue) fits the experimental measurement (lower black dashed line) well. 
The two-component simulation forms a horizontal line at $p=0\,\r{Pa}$ as it neglects excess pore pressure.}
\label{fig:2011time_p}
\end{figure}

The simulation of the loose granular column uses the same parameters as the dense simulation.
The packing density in the column is initialized homogeneously to $\phi = 0.55$ and its height is increased as reported by \cite{rondon2011granular}.
The simulation covering a duration of $6\,\r{s}$ took $213\,\r{h}$ on 8 cores of LEO4.
As a result of the very loose packing, the effective shear strength is low and the column collapses rapidly and entirely, without any static regions.
The pore pressure has to support the majority of the weight and is respectively high (Fig.~\ref{fig:2011l_p}a).
The effective pressure increases at the rapidly flowing front, at $t=0.25\,\r{s}$ (Fig.~\ref{fig:2011l_p}g) due to the dynamic increase of effective pressure following the $\phi(J)$-theory.
The increase in effective pressure leads to a proportional increase in friction and the front is slowed down, Fig.~\ref{fig:2011l_p}h-i.
Although the effective pressure is low in comparison to the dense case (four times lower), the friction is sufficient to bring the slide to a stop.
The final slope inclination is shallow and the low quasi-static particle pressure is sufficient to support the slope, Fig.~\ref{fig:2011l_p}j.
The packing density increases slightly during the collapse but the stability is mostly gained by reducing the slope inclination.
The final pile shape matches the experiment very well, only a small amount of granular material forms a turbidity current that exceeds the runout of the experiment.
The simulated velocity is higher than in the experiment but the difference is less severe than in the dense case.
The simulated excess pore pressure differs remarkably from the measured excess pore pressure as shown in Fig.~\ref{fig:2011time_p}.
Two stages can be observed in the simulated excess pore pressure history.
First, the simulation shows a high peak of excess pore pressure, exceeding the highest experimental pore pressure by a factor of two.
The peak dissipates quickly, as the slide and thus overburden pressure leave the region where the pore pressure sensor is installed.
This first peak is not appearing in the experiment, where the highest pore pressure is reached in a flatter peak at a later point in time.
In a second phase, starting at $t=1\,\r{s}$, the pore pressure dissipates much slower.
In this phase, the pore pressure dissipation is driven by compaction of the granular material and slightly underestimated by the model.

\subsection{Discussion and comparison}
\label{sec:rondon_dc}


The subaqueous granular collapse clearly exceeds the capabilities of the two-component model.
The high sensitivity to the initial packing density can not be explained with this model and the loose and dense simulation are virtually similar.
Results of the two-component model lie between the two extreme cases of the loose and dense experiment, matching the velocity of the loose and the run-out of the dense experiment.
This is reasonable, considering that the missing excess pore pressure is stabilizing the dense column and destabilizing the loose column.
This model is not sufficient for a practical application, as the runout is substantially underestimated in the loose case.
Extremely long run-outs on slopes with $2^\circ$ inclination have been observed in nature \citep[e.g.][]{bryn2005explaining} and they can not be explained with a granular two-component model.

\begin{figure}
\begin{center}
\includegraphics[scale=1]{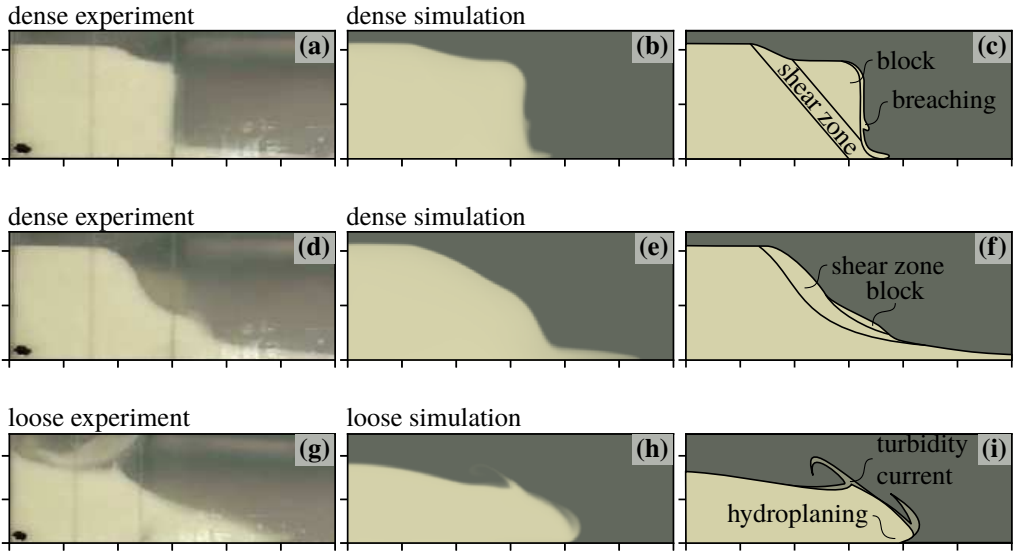}
\end{center}
\caption{Selected snapshots of the experiments from \cite{rondon2011granular} (a,d,g), the simulations (b,e,h) and corresponding sketches (c,f,i). The distance between marks on the axes is $0.02\,\r{m}$. The snapshots highlight the gliding of a cohesive block and breaching (a,b,c), the remoulding of the block due to shearing (d,e,f) and the formation of hydroplaning and turbidity currents (g,h,i) at the loose front.}
\label{fig:sketch}
\end{figure}

The two-phase model can take advantage of its ability to capture excess pore pressure.
It outperforms the two-component model by showing the correct final pile shapes (Figs~\ref{fig:2011d_p}f and \ref{fig:2011l_p}e) and a consistent sensitivity to pore pressure and initial packing density (Fig.~\ref{fig:2011time_p}).
The failure mechanism of both, the dense and the loose experiment, are successfully simulated (see Fig.~\ref{fig:sketch}), indicating that the two-phase model captures the most important physical phenomena.
The model fails in two aspects, as the pore pressure peak in the loose case and the time scale in the dense case differ by a factor of $2$ and $3$, respectively.

It should be noted that no exhausting parameter fitting was required for these results.
Solely the critical state line is optimized to yield the correct pore pressure, all other parameters were selected a priori following \cite{rondon2011granular}, \cite{savage2014modeling}, and \cite{boyer2011unifying}.
Notably, some of the issues, e.g.~the overestimated velocity of the loose collapse, might be resolvable with fitting parameters.
Furthermore, the model allows us to simulate both cases with the same set of parameters with good accuracy.
This distinguishes this work from earlier attempts \citep[e.g.][]{savage2014modeling, wang2017two, si2018development}, where some parameters were fitted individually to the dense and loose case.


Excess pore pressure plays an important role in subaquatic experiments because it controls shear strength and friction.
Dilatancy, compaction, and the dynamic particle pressure further influence friction and thus the kinematics of the slide.
The dense column is only able to collapse after decreasing its packing density and thus its effective shear strength.
The column is dilating until reaching the limiting packing density $\phi_{\r{m}}$.
Before this packing density is reached, the shear rate is limited to the creeping shear rate $S_0$.
A relatively high value was used for this parameter and a lower creeping limit would be desirable, especially considering the error of the time scale in the dense simulation (see appendix~\ref{ssec:creep_shear}).
However, strong oscillations were observed when choosing lower values for $S_0$ because the shear rate often exceeded $S_0$ before dilating sufficiently.

The bottom of the dense column is further compacting in the simulation, up to a packing density of $0.604$.
This is reasonable as the initial particle pressure of $303.3\,\r{Pa}$ at $\phi=0.60$ is below the overburden pressure of $370.8\,\r{Pa}$ of the pile.
At the same time, negative excess pore pressure can be observed at the bottom of the column.
Compaction and negative excess pore pressure seem to contradict each other at first glance.
However, the negative excess pore pressure in the upper parts of the column is so strong, that fluid is flowing upwards from the bottom of the column.
This can be seen in the relative velocity field (Fig.~\ref{fig:2011d_p}h), but also the gradient of pore pressure (Fig.~\ref{fig:2011d_p}b) indicates that pore liquid will flow upwards.

The front speed of the loose collapse is entirely controlled by the dynamic contribution of the $\mu(J)$-$\phi(J)$-rheology to effective pressure and friction.
Simulations with critical state theory (constant friction coefficient $\mu$ and the quasi-static effective pressure model of \cite{johnson1987frictional}) exceed the experimental runout by far (see Appendix~\ref{ssec:dynamics}).
This is a strong contrast to the subaerial case where acceptable results could be achieved with critical state theory.

The dynamic contribution to particle pressure and friction plays also an important role in the dense case, although this pile collapses very slow.
The thin layers of grains that are breaching from the unsupported column flank reach packing densities far below $\phi_{\r{rlp}} = 0.53$ due to mixing with the ambient fluid.
At this packing density, the quasi-static contribution to effective pressure vanishes, and the runout of these particles is entirely controlled by dynamic particle pressure and friction.
The runout of the breaching flank could not be controlled in simulations with critical state theory (see appendix~\ref{ssec:dynamics}).

The pore pressure in the loose case differs qualitatively and quantitatively from the measurement.
Within the applied model, it seems reasonable that a high initial peak decreases quickly, as substantial amounts of grains and thus overburden pressure leave the region of the pressure sensor.
Similar results with an early, short, and strong peak and a slow further dissipation, close to the measurement, have been obtained with other frameworks, e.g.~by \cite{bouchut2017two} or \cite{baumgarten2019general}.


The dilatancy of the dense column is substantially faster in the numerical model than in the experiment, although the permeability is underestimated following the comparison of the pore pressure.
Therefore it is unlikely that permeability is the cause for this discrepancy and we assume that inaccuracies in the rheology are responsible.
The $\mu(J)$-$\phi(J)$-rheology describes the steady shearing of a fluid-grain mixture very well \citep{boyer2011unifying}.
However, the transient transition towards the steady packing density at a certain effective pressure is not described.
This transition depends on the permeability of the granular material but also on its viscosity (shear and bulk viscosity).
As mentioned before, the high value for the creeping shear rate $S_0$ could be responsible for this issue but it might also be related to the missing bulk viscosity or a mismatch of constitutive parameters.
Bulk viscosity could delay the dilatancy in the early stage of the dense collapse, bringing the time scale of the collapse in the simulation closer to the experiment.
Bulk viscosity could further help to decrease the pore pressure peak in the loose case, as some of the pore pressure could be transformed into viscous pressure.
\cite{schaeffer2019constitutive} suggests a form for the bulk viscosity which has the potential to improve these aspects.


\cite{savage2014modeling} and \cite{si2018development} include a cohesive shear strength into their model to correct some of these problems and to fit results to the experiment.
However, there is no evidence for cohesive forces in a fully submerged granular flow.
Neither electrostatic forces nor cementing have been reported by \cite{rondon2011granular}.
Apparent cohesion can be traced back to negative excess pore pressure, which is directly simulated by the numerical model.
Notably, \cite{si2018development} are able to control the slide velocity very well.
However, this is achieved by fitting the cohesion to the respective case and by a strong overestimation of the negative excess pore pressure, reaching values around $500\,\r{Pa}$ at the pressure sensor at $t=3\,\r{s}$ \cite[see Fig.~5 by][]{si2018development}).

\cite{baumgarten2019general} applied a similar model (elasto-plastic multiphase model with $\mu(K)$-$\phi(K)$-scaling) to the same cases.
The results show similar problems, i.e. an overestimation of the pore pressure in the loose case and an overestimation of the collapse velocity in the dense case.
Notably, we achieve similar results in these test cases with a substantially simpler model.


\section{Conclusions}
\label{sec:conclusion}


The Navier-Stokes Equations can be an adequate tool for accurate simulations of granular flows when they are complemented with the correct rheologies.
Substantial progress has been made in recent years with the $\mu(I)$-rheology and its extensions to compressible flows and low Stokes number flows.
The incompressible $\mu(I)$-rheology fits well into the multi-component framework of OpenFOAM and the compressible $\mu(I)$-$\phi(I)$-rheology fits well into the multi-phase framework, as previously shown by e.g.~\cite{chauchat2017sedfoam}.
We apply, for the first time, the compressible $\mu(I)$-$\phi(I)$-rheology to granular collapses and avalanching flows.
The superposition with the critical state theory is imperative to get realistic packing densities at rest and a stable solver.
For subaerial, i.e.~high Stokes number flows, dilatancy plays a minor role and results of the compressible model are similar to the incompressible model.
However, the dilatancy predicted by the compressible model is able to close the gap between the experiments and the incompressible model.
Further, the compressible model should be well-posed \citep{barker2017well, heyman2017compressibility, schaeffer2019constitutive}, in contrast to many incompressible granular flow models \citep{barker2015well}.
Note that bulk viscosity, which is imperative for a well-posed rheology \citep[e.g.][]{schaeffer2019constitutive}, was not considered in this study.
However, the coupling of the granular phase to the pore fluid has a similar effect as bulk viscosity and might be able to restore a well-posed system.
For a guaranteed well-posed compressible rheology that collapses to the $\mu(I)$-$\phi(I)$-rheology in steady state, the reader is referred to \cite{schaeffer2019constitutive}.

The upsides of the compressible two-phase model come at the cost of more parameters and a stronger mesh dependence.
Furthermore, code and case setup are more complicated with the two-phase model and simulations are more prone to failure if initial conditions or parameters are not well suited for the case.
Therefore, the incompressible model might be better suited for some flows at high Stokes numbers, especially considering regularized rheologies that are well-posed for a wide range of flow regimes \citep[e.g][]{barker2017partial}.
Notably, we did not encounter any problems with the partial ill-posedness of the $\mu(I)$-rheology, which could be related to relatively coarse grids, high numerical diffusion, the short simulation duration or the truncation of the viscosity.


The extension to low Stokes number flows is made possible by the $\mu(J)$-$\phi(J)$-rheology.
At low Stokes numbers, it is imperative to consider excess pore pressure and a two-phase model is required.
Therefore, the incompressible $\mu(J)$-rheology is rather impractical and becomes only applicable after supplementing it with the $\phi(J)$-curve to the compressible $\mu(J)$-$\phi(J)$-rheology. 
The dynamic growth of pressure and friction is substantial for accurate results, highlighting the value of the $\mu(J)$-$\phi(J)$-rheology.
The fitting of parameters was reduced to a minimum and only the critical state line had to be optimized to the experiments.
It should be noted that these parameters could be determined by measuring the critical packing density at a few pressure levels, making the simulations free of any fitted parameter.
The compressible two-phase model reacts sensitive to the packing density, recreating the final runout, pile shape, and failure mechanism of the experiments very well.
The model still lacks in some aspects, e.g.~the time scale and the velocity of the dense collapse and the pore pressure peak in the loose collapse.


It was shown that the incompressible two-component model can be derived from the compressible two-phase model by neglecting the relative velocity between phases.
This simplification yields reasonable results for subaerial granular flows at high Stokes numbers but fails to describe the subaquatic granular flows at low Stokes numbers.
This seems to be contradictory, as the relative velocity (which was neglected in the incompressible model) is very small in the subaquatic case (see Figs.~\ref{fig:2011d_p} and \ref{fig:2011l_p}) but considerable high in the subaerial case (see Fig.~\ref{fig:2005_tp_p}).
This apparent paradox can be resolved by the fact that unhindered density changes have no notable influence on the flow dynamics.
However, if changes in packing density are constrained, pore pressure will build up and the rheology of the material will change drastically.
Thus, pore pressure, rather than compressibility is the key factor that allows the two-phase model to accurately capture the flow mechanics.
The two-phase model provides many other upsides aside from the inclusion of pore pressure.
The continuous transition from dense granular material to pure ambient fluid should be useful for the simulation of granular free streams \citep[][]{viroulet2017multiple}, turbidity currents \citep{heerema2020determines} and powder snow avalanches \citep[e.g.][]{sovilla2015structure}.
Other studies showed that the two-phase model is useful for sediment transport \citep{chauchat2017sedfoam} and other dilute particle-fluid mixtures \citep[e.g.][]{passalacqua2011implementation}.


OpenFOAM provides a good platform to evaluate concepts (e.g.~the multi-component and multi-phase methodology) and models (e.g.~$\mu(I)$-$\phi(I)$ and $\mu(J)$-$\phi(J)$-rheologies).
The implemented rheologies can be further coupled with segregation \citep{barker2020coupling} or tsunami simulations \cite[e.g.][]{si2018general}.
However, the segregated semi-implicit solver strategy of OpenFOAM sets limits to models and execution velocity, as (part of the) viscous terms and the particle pressure are included explicitly.
This showed to be problematic and a fully implicit solver, that solves all equations simultaneously, might be superior in this regard.


The model can help to understand the extreme outruns of submarine landslides, such as the Storegga landslide \citep[e.g.][]{bryn2005explaining} and the big variation in tsunamigenic potentials \citep[e.g.][]{lovholt2017some}.
Theories, such as hydroplaning and remoulding \citep[e.g.][]{de2004hydroplaning} can be quantitatively described by the critical state theory and its dynamic extension in form of the $\mu(J)$-$\phi(J)$-rheology.
Hydroplaning, formerly described as the flowing of sediment on a thin layer of liquid can be interpreted as a region of low or even zero packing density and vanishing effective pressure.
This can be observed in Fig.~\ref{fig:sketch}g-i, where the front of the loose slide is lifted by pressure in the surrounding fluid.
Remoulding can similarly be explained with critical state theory as an overconsolidated sample that is dilating during shearing (see Fig.~\ref{fig:sketch}a-f).
The two-phase model and its capability to describe various and realistic failure mechanisms with different time scales are particularly valuable for understanding the tsunamigenic potential of submarine landslides and the respective slopes.
The dense column collapses very slowly, reaching velocities of up to $0.1\,\r{m\,s^{-1}}$ in small layers near the surface.
The loose column collapses entirely with velocities up to $0.4\,\r{m\,s^{-1}}$.
The tsunamigenic potential of a landslide scales with initial acceleration and the mobilized volume \citep[e.g.][]{lovholt2017some} and a substantial difference in tsunamigenic potential follows for the dense and the loose slide.
This shows that packing density, excess pore pressure and permeability are key parameters in controlling stability, failure mechanism, slide acceleration, and tsunamigenic potential.

Many full-scale subaquatic landslide simulations are based on Bingham fluids, a visco-plastic rheology independent of the pressure \citep[e.g.][]{kim2019landslide}.
This seems to stand in strong contradiction to the model applied here.
However, the simulation of the loose case shows that packing density changes are small.
For a nearly constant packing density, the effective pressure decouples from overburden pressure because the weight is absorbed entirely by pore pressure.
As a consequence, overburden pressure and friction will decouple and the microscopic granular friction will appear as cohesion on a macroscopic scale.
The macroscopic description as a Bingham fluid is therefore surprisingly consistent with the findings in this work, especially for fine grained marine sediments with low permeabilities.

\section{Summary}
\label{sec:summary}

This work highlights a path to extend the incompressible $\mu(I)$-rheology for subaerial granular flows to the compressible $\mu(J)$-$\phi(J)$-rheology for subaquatic granular flows.
The implementation of the $\mu(I)$-$\phi(I)$-rheology in a multiphase framework and the $\mu(I)$-rheology in a multi-component framework allows us to conduct subaerial granular collapses with two different models.
The application shows consistency between the incompressible $\mu(I)$-rheology \citep[e.g.][]{lagree2011granular} and the compressible $\mu(I)$-$\phi(I)$-rheology.
Notably, substantial modifications to the $\phi(I)$-curve are required for a practical application of the rheology.
The simulations show that compressibility and dilatancy have a small influence on high Stokes number flows because excess pore pressure is negligibly small.

The implementation of the $\mu(J)$-$\phi(J)$-rheology extends possible applications to low Stokes number flows, e.g.~subaquatic granular collapses.
The incompressible model reaches its limitations under these conditions and the compressible model is required for an accurate simulation.
Other than previous attempts, we applied the exact same set of parameters to an initially dense and loose granular collapse with satisfying results.
Notably, the application of the $\mu(J)$-$\phi(J)$-rheology does not require an extensive fitting of constitutive parameters. 
The comparison between the compressible model and experiments uncovered discrepancies in the time scale and the pore pressure.
These could be indicators for issues in the rheology, e.g.~a missing bulk viscosity or issues with the creeping regime that had to be introduced for numerical stability.
The well-posedness of the proposed model is not guaranteed and should be investigated in the future.

The compressible two-phase model has a wide range of applications and the results have implications on many problems in geoscience.
Applications to sediment transport and scouring \citep{cheng2017sedfoam} have been shown with a similar model.
We further expect the applicability to turbidity currents and all other gravitational mass flows with low and high Stokes numbers.
Furthermore, \cite{si2018general} showed the applicability of a similar model to landslide tsunami simulations by incorporating the free water surface.

\backsection[Acknowledgements]{The author gratefully acknowledges the support from Wolfgang Fellin and the research area scientific computing of the University of Innsbruck. The author thanks Gertraud Medicus, Thomas Barker, Finn L{\o}vholt and Geir Pedersen for helpful comments and Pascale Aussillous for authorizing the use of pictures of the experiment. Further, the author thanks the editor and two referees for their helpful comments and support in publishing this work.}

\backsection[Funding]{This project has received funding from the European Union's Horizon 2020 research and innovation programme under the Marie Sk{\l}odowska-Curie grant agreement No.~721403 (SLATE). The computational results presented have been achieved (in part) using the HPC infrastructure LEO of the University of Innsbruck.}

\backsection[Declaration of interests]{The author reports no conflict of interest.}

\backsection[Author ORCID]{M. Rauter, https://orcid.org/0000-0001-7829-6751}

\appendix

\section{Derivation of the two-component model}
\label{sec:derivation}

The two-component model can be derived from the two-phase model by summing up the mass and momentum conservation equations.
The sum of the mass conservation equations \eqref{eq:disp_alpha} and \eqref{eq:disp_alpha_s} yields
\begin{equation}
\dfrac{\partial \phi_{\r{c}} + \phi_{\r{g}}}{\partial t} + \bnabla\bcdot\left(\phi_{\r{c}}\,\b{u}_{\r{c}} + \phi_{\r{g}}\,\b{u}_{\r{g}}\right) = 0
\end{equation}
and with the definitions $\phi_{\r{c}} + \phi_{\r{g}} = 1$ and $\ol{\b{u}} = \phi_{\r{c}}\,\b{u}_{\r{c}} + \phi_{\r{g}}\,\b{u}_{\r{g}}$ we can derive the continuity equation of the two-component model, Eq.~\eqref{eq:NS_cont}.

The sum of the momentum conservation equations \eqref{eq:disp_momentum} and \eqref{eq:disp_momentum_s} is slightly more complex and approximations are required due to non-linearities.
Therefore, we will cover each term individually in the following.
The sum of the time derivatives of Eqs.~\eqref{eq:disp_momentum} and \eqref{eq:disp_momentum_s} can be simplified with the definition of the volume averaged velocity $\ol{\b{u}}$, the local density $\rho = \phi_{\r{c}}\,\rho_{\r{c}} + \phi_{\r{g}}\,\rho_{\r{g}}$ and the relative velocity $\b{u}_{\r{r}} = \b{u}_{\r{g}} - \b{u}_{\r{c}}$ to 
\begin{eqnarray}
&\dfrac{\partial}{\partial t} \left(\phi_{\r{c}}\,\rho_{\r{c}}\,\b{u}_{\r{c}} + \phi_{\r{g}}\,\rho_{\r{g}}\,\b{u}_{\r{g}}\right) = 
\dfrac{\partial}{\partial t} \left(\phi_{\r{c}}\,\rho_{\r{c}}\,\left(\ol{\b{u}}-\phi_{\r{g}}\b{u}_{\r{r}}\right) + \phi_{\r{g}}\,\rho_{\r{g}}\,\left(\ol{\b{u}}+\phi_{\r{c}}\,\b{u}_{\r{r}}\right)\right) =\nonumber\\
&\dfrac{\partial \rho\,\ol{\b{u}}}{\partial t} + \dfrac{\partial }{\partial t} \left(\phi_{\r{g}}\,\phi_{\r{c}}\,\b{u}_{\r{r}}\,\left(\rho_{\r{g}}-\rho_{\r{c}}\right)\right) \approx \dfrac{\partial \rho\,\ol{\b{u}}}{\partial t}.\label{eq:time_derivative}
\end{eqnarray}
The second term in Eq.~\eqref{eq:time_derivative} vanishes if the relative velocity is zero or if the phase densities are equal.
The two-component model assumes that phase velocities are equal (Eq.~\ref{eq:samevelocities}) and the second term can be neglected.
The error in the momentum conservation is expected to be small in relation to limitations of the incompressible rheology.

The sum of the convective fluxes follows a similar pattern,
\begin{eqnarray}
&\bnabla\bcdot\left(\phi_{\r{c}}\,\rho_{\r{c}}\,\b{u}_{\r{c}}\otimes\b{u}_{\r{c}}+\phi_{\r{g}}\,\rho_{\r{g}}\,\b{u}_{\r{g}}\otimes\b{u}_{\r{g}}\right) = \nonumber \\
&\bnabla\bcdot\left(\phi_{\r{c}}\,\rho_{\r{c}}\,\left(\ol{\b{u}}-\phi_{\r{g}}\,\b{u}_{\r{r}}\right)\otimes\left(\ol{\b{u}}-\phi_{\r{g}}\,\b{u}_{\r{r}}\right)+\phi_{\r{g}}\,\rho_{\r{g}}\,\left(\ol{\b{u}}+\phi_{\r{c}}\,\b{u}_{\r{r}}\right)\otimes\left(\ol{\b{u}}+\phi_{\r{c}}\,\b{u}_{\r{r}}\right)\right) = \nonumber \\
&\bnabla\bcdot\left(\rho\,\ol{\b{u}}\otimes\ol{\b{u}}\right) +
2\bnabla\bcdot\left(\phi_{\r{c}}\,\phi_{\r{g}}\,\ol{\b{u}}\otimes\b{u}_{\r{r}}\left(\rho_{\r{g}}-\rho_{\r{c}}\right)\right) +
\bnabla\bcdot\left(\phi_{\r{c}}\,\phi_{\r{g}}\,\b{u}_{\r{r}}\otimes\b{u}_{\r{r}}\left(\phi_{\r{g}}\,\rho_{\r{c}}+\phi_{\r{c}}\,\rho_{\r{g}}\right)\right) \approx \bnabla\bcdot\left(\rho\,\ol{\b{u}}\otimes\ol{\b{u}}\right).
\end{eqnarray}
The second and third term vanish if the relative velocity is zero.
Notably, only the second term vanishes if the phase densities are equal due to the non-linearity in the convection term.
For the approximation of the two-component model, it is sufficient to recover the first term, as the relative velocity is neglected.

The terms on the left hand side of Eqs.~\ref{eq:disp_momentum} and \ref{eq:disp_momentum_s}, can be summed up without further assumptions,
\begin{eqnarray}
&\bnabla\bcdot\left(\phi_{\r{c}}\,\bt{T}_{\r{c}} + \phi_{\r{g}}\,\bt{T}_{\r{g}} \right) = \bnabla\bcdot\bt{T},\\
&\phi_{\r{c}}\,\bnabla\,p+\phi_{\r{g}}\,\bnabla\,p + \bnabla\,p_{\r{s}} = \bnabla p_{\r{tot}},\\
&\phi_{\r{c}}\,\rho_{\r{c}}\,\b{g}+\phi_{\r{g}}\,\rho_{\r{g}}\,\b{g} = \rho\,\b{g},\\
&k_{\r{gc}} \left(\b{u}_{\r{g}}-\b{u}_{\r{c}}\right) + k_{\r{gc}} \left(\b{u}_{\r{c}}-\b{u}_{\r{g}}\right) = 0,
\end{eqnarray}
and the momentum-conservation equation of the two-component model, Eq.~\eqref{eq:NS_momentum} can be assembled.

\section{Sensitivity study}
\label{sec:convergence}

The numerical models require a wide range of parameters.
Most parameters are physical and can be derived from experiments and literature.
However, some parameters are purely numerical and their values can not be derived from experiments.
The following parameter study was used to derive numerical parameters that have been applied in simulations.
The most influential numerical parameters are the grid size $\Delta x$, the Courant number $\r{CFL}^{\r{diff}}$, related to the time step $\Delta t$, and the maximum viscosity $\nu_{\max}$.
Furthermore, the effect of dynamic friction, i.e.~the difference between a constant friction coefficient $\mu$ and the $\mu(I)$ or $\mu(J)$-rheology is investigated.
The applied model and the flow regime (subaerial or subaquatic, dense or loose) are shown for each figure in the upper left edge.

\subsection{Influence of the maximum viscosity}
\label{ssec:convergence_viscosity}

The maximum viscosity is one of the most influential numerical parameters in the applied model.
It should be reasonably high to mimic solid behaviour, but as small as possible to improve numerical stability and to keep computational expense low.
A reasonable limit can be found by investigating the dimensionless governing equations, in which the respective scales are isolated.
The momentum conservation of the Navier-Stokes Equations can be written as 
\begin{equation}
\dfrac{\partial \b{u}}{\partial t} + \bs{\nabla}\bcdot\left(\b{u}\otimes\b{u}\right) = -\dfrac{1}{\rho}\,\bs{\nabla}\,p + \dfrac{1}{2}\,\nu\,\bs{\nabla}^2\,\b{u} + \b{g}\label{eq:momentum_simple}
\end{equation}
for a single incompressible Newtonian fluid with density $\rho$ and constant viscosity $\nu$.
By scaling space with the height of the slide $H$, the velocity with the respective free fall velocity $\sqrt{|\b{g}|\,H}$, the time with the free fall time $\sqrt{H/|\b{g}|}$, and the pressure with the respective hydrostatic pressure $\rho\,|\b{g}|\,H$, the dimensionless variables (marked with a hat) can be established as
\begin{eqnarray}
&\b{x} = H\,\hat{\b{x}},\\
&\bs{\nabla} = \dfrac{1}{H}\,\hat{\bs{\nabla}} = \dfrac{1}{H}\,\left(\dfrac{\partial}{\partial\hat{\b{x}}}, \dfrac{\partial}{\partial\hat{\b{y}}}, \dfrac{\partial}{\partial\hat{\b{z}}}\right)^T,\\
&\b{u} = \sqrt{|\b{g}|\,H}\,\hat{\b{u}},\\
&t = \sqrt{\dfrac{H}{|\b{g}|}}\,\hat{t},\\
&p = \rho\,|\b{g}|\,H\,\hat{p}.
\end{eqnarray}
Introducing the dimensionless variables into the momentum conservation equation and dividing by $|\b{g}|$ yields
\begin{equation}
\dfrac{\partial \hat{\b{u}}}{\partial \hat{t}} + \hat{\bs{\nabla}}\bcdot\left(\hat{\b{u}}\otimes\hat{\b{u}}\right) = -\hat{\bs{\nabla}}\,\hat{p} + \sqrt{\dfrac{1}{|\b{g}|\,H^3}} \nu\,\hat{\bs{\nabla}}^2\,\hat{\b{u}} + \dfrac{\b{g}}{|\b{g}|}.
\end{equation}
In the case of a solid-like behaviour, the viscous term should be dominating over all other terms.
All terms, except for the viscous term, are of order one and we can deduce that the inequality
\begin{equation}
\sqrt{\dfrac{1}{|\b{g}|\,H^3}}\nu_{\r{max}} > \dfrac{1}{\varepsilon}
\end{equation}
should be fulfilled to simulate the behaviour of a solid.
$\varepsilon$ is a small dimensionless number, indicating the magnitude of viscous stresses over other terms.
The viscosity which is required for a solid-like behaviour can be calculated as 
\begin{equation}
\nu_{\r{max}} > \dfrac{1}{\varepsilon}\,\sqrt{|\b{g}|\,H^3},
\end{equation}
as a function of respective scales by choosing the magnitude of viscous stresses over other terms, $\varepsilon$.
The required magnitude for $\varepsilon$ can be estimated by conducting a numerical sensitivity analysis.

\begin{figure}
\begin{center}
\includegraphics[scale=1]{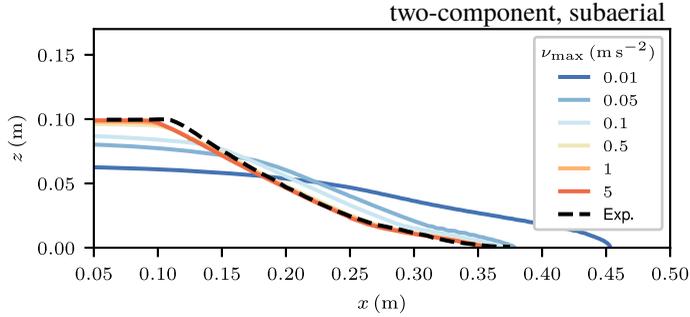}
\end{center}
\caption{Pile shape at $t=0.8\,\r{s}$ of the subaerial granular collapse with various values for $\nu_{\r{max}}$ using the two-component model. The high influence of this numerical parameter and the unphysical effect of low values is clearly visible. The dashed black line shows the final pile shape of the experiment for comparison. The two-phase model behaves similarly.}
\label{fig:2005_nu}
\end{figure}

Variations of $\nu_{\r{max}}$ (and thus $\varepsilon$) are presented in Fig.~\ref{fig:2005_nu} for the subaerial case at $t=0.8\,\r{s}$, using the two-component model.
The value of $\nu_{\r{max}} = 1\,\r{m^{2}\,s^{-1}}$ is adequate for this example and the left side of the pile stays nearly static as it is the case in the experiment.
The respective value for the dimensionless scaling factor $\varepsilon$ follows as $0.1$ ($H = 0.1\,\r{m}$, $|\b{g}|\approx 10\,\r{m\,s^{2}}$), indicating that viscous forces have to be about $10$ times higher than all other contributions to the momentum conservation equation.
For lower viscosities, the pile is notably deformed and shows no stable regions and no granular characteristics.
Rather, the pile shows the characteristics of a visco-plastic fluid \citep[see rheology comparisons by][]{lagree2011granular}, which indicates that the viscosity threshold was dominating the simulation.
Notably, cases with high maximum viscosity are stable after $t=0.4\,\r{s}$ while cases with low maximum viscosity keep flowing beyond $t=0.8\,\r{s}$.
For an application of granular rheologies in OpenFOAM, we suggest a maximum viscosity following Eq.~\ref{eq:nu_thresh} with $\varepsilon = 1/10$.

\subsection{Grid sensitivity}
\label{ssec:convergence_grid}

\begin{figure}
\begin{center}
\includegraphics[scale=1]{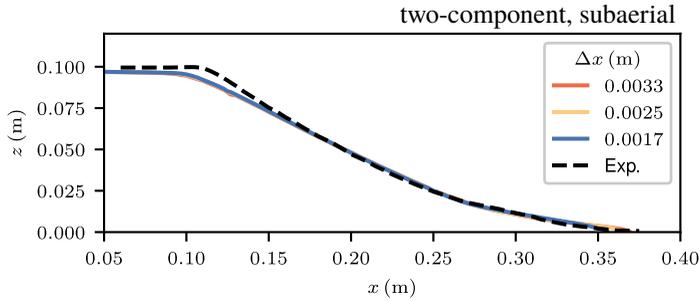}
\end{center}
\caption{Grid sensitivity of the two-component model for the subaerial case. The model behaves similarly in the subaquatic cases. The black dashed line shows the experimental final pile shape for reference.}
\label{fig:singlephase_dx}
\end{figure}

The grid sensitivity is an important issue for complex flow models and we provide a full grid sensitivity analysis for the multi-component and the multi-phase model.
The grid sensitivity study for the multi-component model is solely conducted for the subaerial case because the mechanics of this model is similar in all cases.
The final pile shape of the investigated case is shown in Fig.~\ref{fig:singlephase_dx} for various grid resolutions.
This model reacts very robustly to coarse grids and 30 cells along the pile height are sufficient to get accurate results for the final pile shape.

\begin{figure}
\begin{center}
\includegraphics[scale=1]{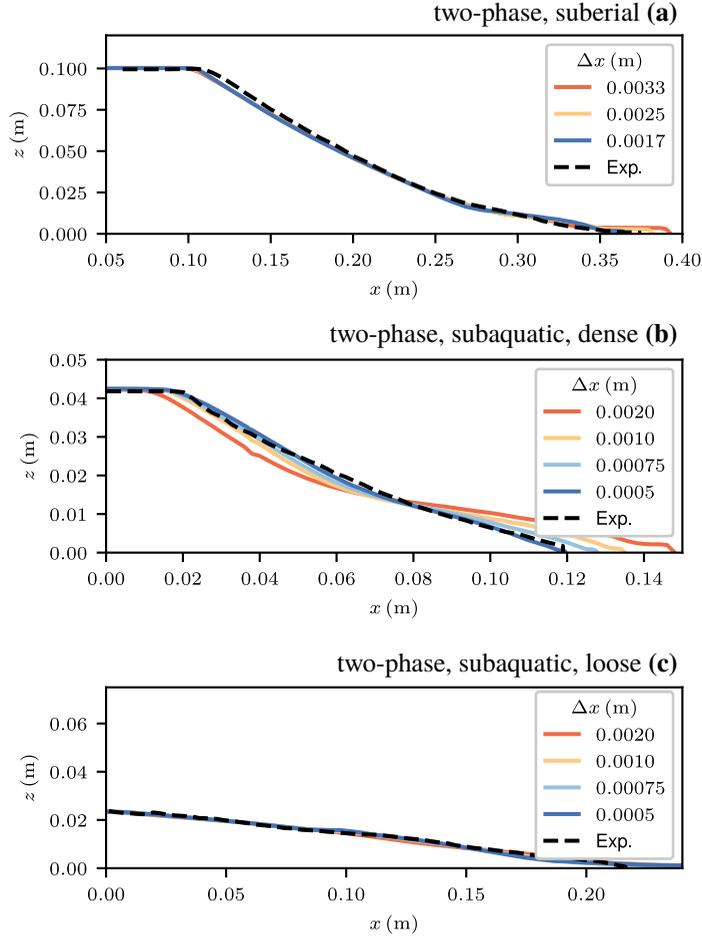}
\end{center}
\caption{Grid sensitivity of the two-phase model for the subaerial case (a), the dense subaquatic case (b) and the subaquatic loose case (c). The black dashed lines show the experimental final pile shapes for reference.}
\label{fig:twophase_dx}
\end{figure}

The two-phase model is more complex in terms of grid sensitivity.
Three different failure mechanisms of the granular column can be observed in the simulations with the two-phase model.
The three mechanisms react differently to a variation of the grid resolution.
In some cases, the model reacts very sensitively to coarse grids and a grid refinement study should always be performed when applying this model.
Fig.~\ref{fig:twophase_dx} shows the grid convergence analysis for the subaerial, the subaquatic dense, and subaquatic loose case.
The two-phase model is slightly more sensitive to coarse girds than the two-component model in the subaerial case, see Fig.~\ref{fig:twophase_dx}a.
The problematic area is the thin flow front and the issue is probably related to the mixed role of the phase-fraction field.

The two-phase model is very sensitive to coarse grids in the subaquatic dense case, see Fig.~\ref{fig:twophase_dx}b.
Breaching of a thin layer of grains on the unsupported side of the column leads to a reduced phase-fraction in cells that contain the slide surface.
This reduces the effective pressure in all of those cells and further the shear strength, accelerating the collapse.
The result is a mesh dependency of the final pile shape and the collapse velocity.
The mesh had to be refined down to a cell size of $0.0005\,\r{m}$, to achieve accurate results.
Notably, the difference between the smallest two cell sizes is still remarkably at the front of the collapse.
An additional refinement step would be desirable, but this would have exhausted the available computational resources.

The loose subaquatic case can be simulated with good accuracy on a relatively coarse mesh as shown in Fig.~\ref{fig:twophase_dx}c.
This is not surprising as the failure mechanism and flow pattern is much simpler.
In particular, the effective pressure discontinuity at the free surface is weaker than in the dense case and thus requires a smaller grid resolution.

\subsection{Time step duration sensitivity}
\label{ssec:convergence_timestep}

The time step duration is investigated similarly as the grid resolution.
The time step duration is not fixed but adapted to velocity ($\r{CFL}^{\r{conv}}$) and viscosity ($\r{CFL}^{\r{diff}}$), relative to the grid size.
The viscous contribution to the $\r{CFL}$ number is always much bigger in the investigated cases and the time step sensitivity study was conducted based on this value.
Notably, stability can be guaranteed only for $\r{CFL}^{\r{diff}} <1$.
However, the $\r{CFL}^{\r{diff}}$ number allows only an estimation of the stability and in some cases larger time steps are possible.

The two-component solver can operate in two modes, using a full momentum predictor step or a reduced momentum predictor step.
The full momentum predictor step solves the full linearised system of the discretized momentum conservation equation.
The reduced momentum predictor step calculates an explicit prediction of the velocity field based on the velocity field of the last time step.
This has a substantial influence on the stability when viscous stresses are dominating.

\begin{figure}
\begin{center}
\includegraphics[scale=1]{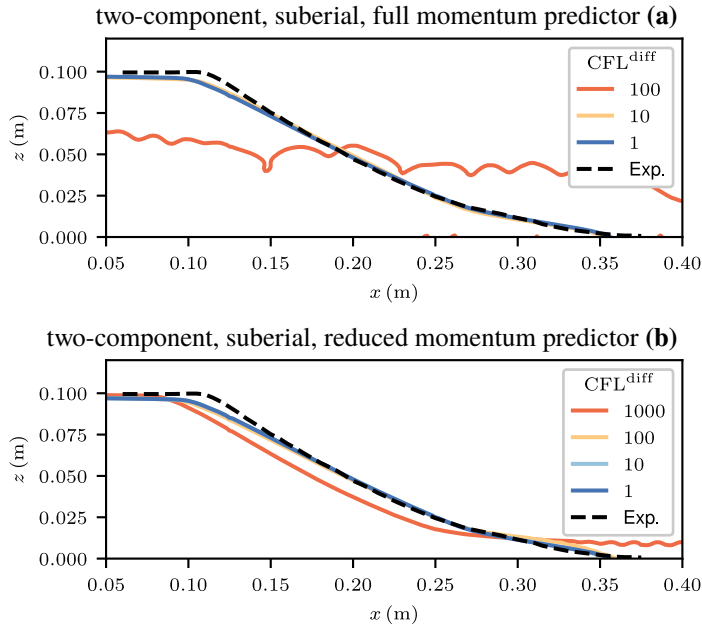}
\end{center}
\caption{Sensitivity of the two-component model on the time step duration, expressed by the viscous CFL number. The solver was operated with the full momentum predictor (a) and the reduced momentum predictor (b).}
\label{fig:singlephase_dt}
\end{figure}

Fig.~\ref{fig:singlephase_dt}a shows the final pile shape for various time step durations and the full momentum predictor step in the two-component model.
Notably oscillations in pressure can already be seen at $\r{CFL}^{\r{diff}} = 2$ (not shown) and they grow substantially for higher $\r{CFL}^{\r{diff}}$ numbers.
The pressure oscillations start to influence the pile shape at $\r{CFL}^{\r{diff}} = 10$ and the pile is completely distorted for $\r{CFL}^{\r{diff}} = 100$.

The two-component model is more robust to larger time steps when operating with the reduced momentum predictor step, see Fig.~\ref{fig:singlephase_dt}b.
Pressure oscillations start approximately at $\r{CFL}^{\r{diff}} = 100$ and the first influence on the slide geometry can be observed at $\r{CFL}^{\r{diff}} = 1000$.
Anyway, it is recommended to run the two-component model with small enough time steps to prevent pressure oscillations, ideally at $\r{CFL}^{\r{diff}} = 1$, as done in this work.

\begin{figure}
\begin{center}
\includegraphics[scale=1]{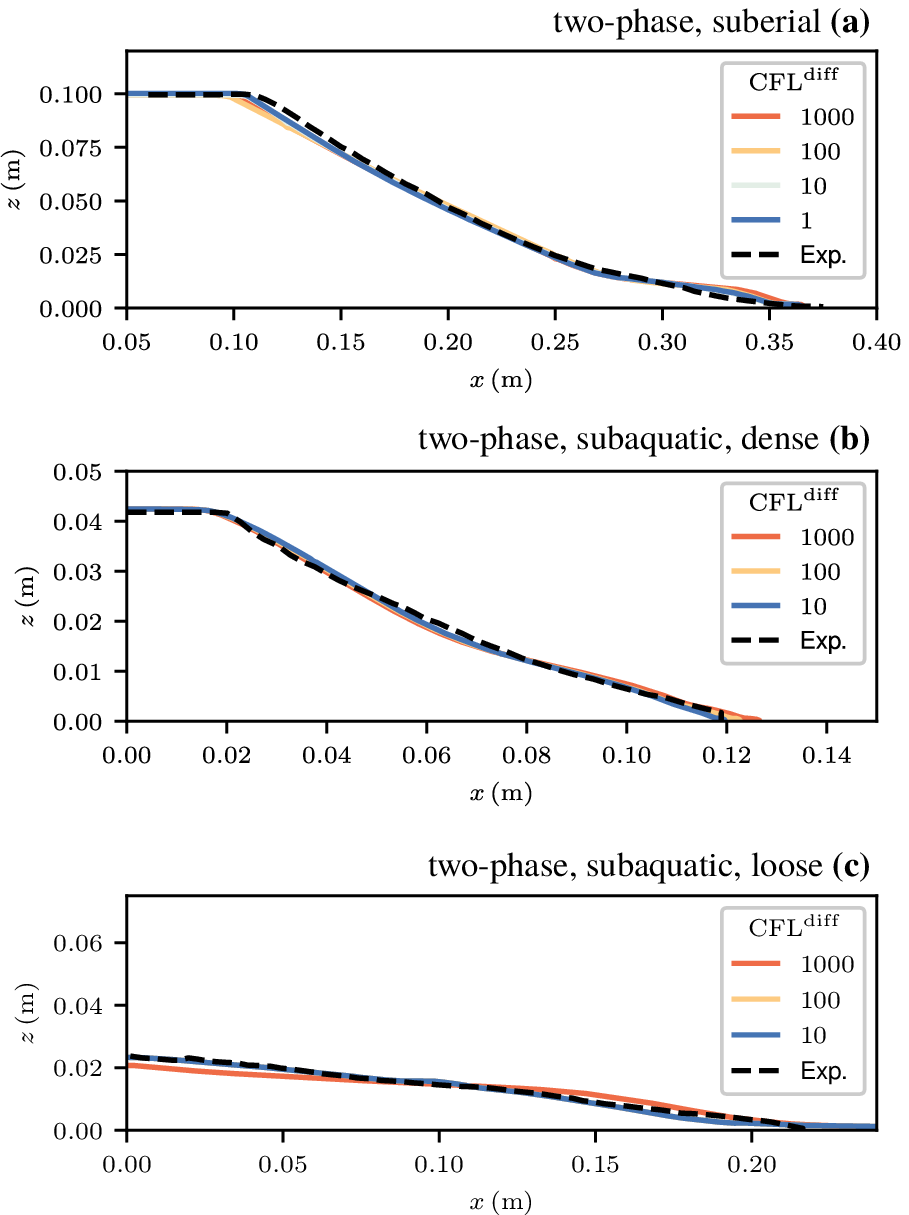}
\end{center}
\caption{Sensitivity of the two-phase model on the time step duration, expressed by the viscous CFL in the subaerial case (a), the subaquatic, dense case (b) and the subaquatic, loose case (c).}
\label{fig:twophase_dt}
\end{figure}

The two-phase model reacts less sensitive to large time step durations.
In fact, simulations were stable up to $\r{CFL}^{\r{diff}} = 1000$.
No pressure oscillations could be observed and the final pile shape is nearly unaffected for all cases, see Fig.~\ref{fig:twophase_dt}.
However, the accuracy got worse for large time step durations and we observed a slower initial acceleration for $\r{CFL}^{\r{diff}} = 1000$.
No subaquatic simulations with the two-phase model and $\r{CFL}^{\r{diff}} = 1$ have been conducted, as this would have exhausted the computational resources.

\subsection{Influence of dynamic friction and wall friction}
\label{ssec:dynamics}

\begin{figure}
\begin{center}
\includegraphics[scale=1]{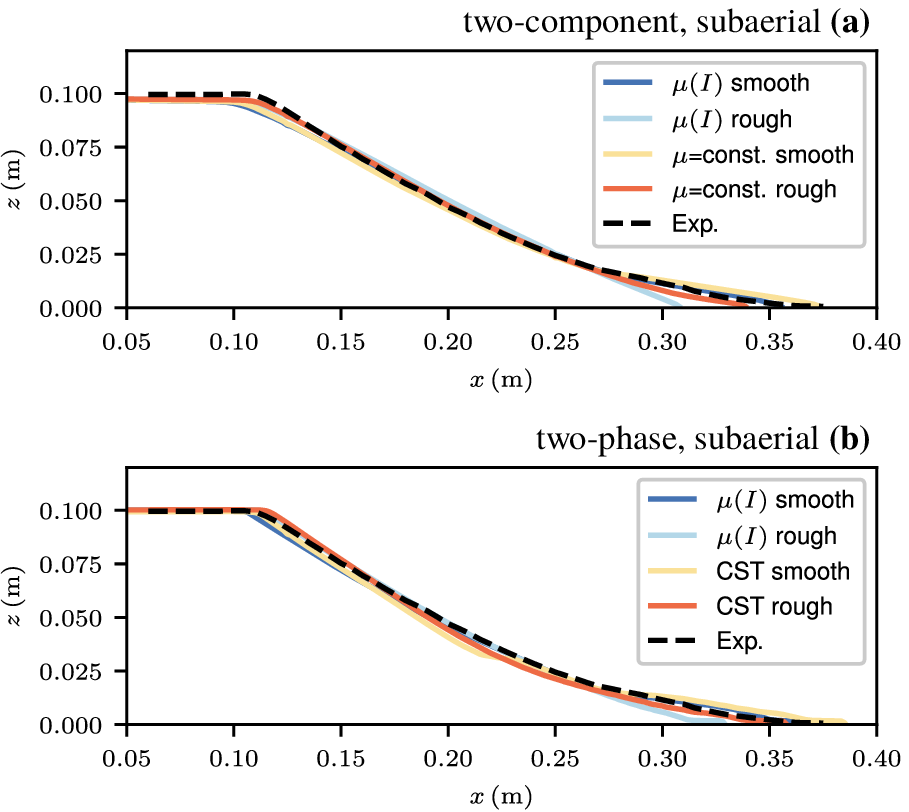}
\end{center}
\caption{Influence of dynamic friction, dynamic effective pressure and wall friction on the final pile shape in the two-component model (a) and the two-phase model (b).}
\label{fig:2005_dynamics}
\end{figure}

The effect of the dynamic increase of friction following the $\mu(I)$-rheology and the effect of the reduced wall friction were investigated with both models.
Results are shown in Fig.~\ref{fig:2005_dynamics}a for the two-component model and in Fig.~\ref{fig:2005_dynamics}b for the two-phase model.
The models do not react sensitively to the variation of the friction model and the basal friction coefficient.
The runout is slightly underestimated in simulations with the $\mu(I)$-rheology on a rough surface.
However, the introduction of the smooth surface elongates the runout and the simulations fit the experiment well.
Simulations with a constant friction coefficient on a rough surface fit the experiment also well, simulations with constant friction coefficient on a smooth surface overestimate the runout.

\begin{figure}
\begin{center}
\includegraphics[scale=1]{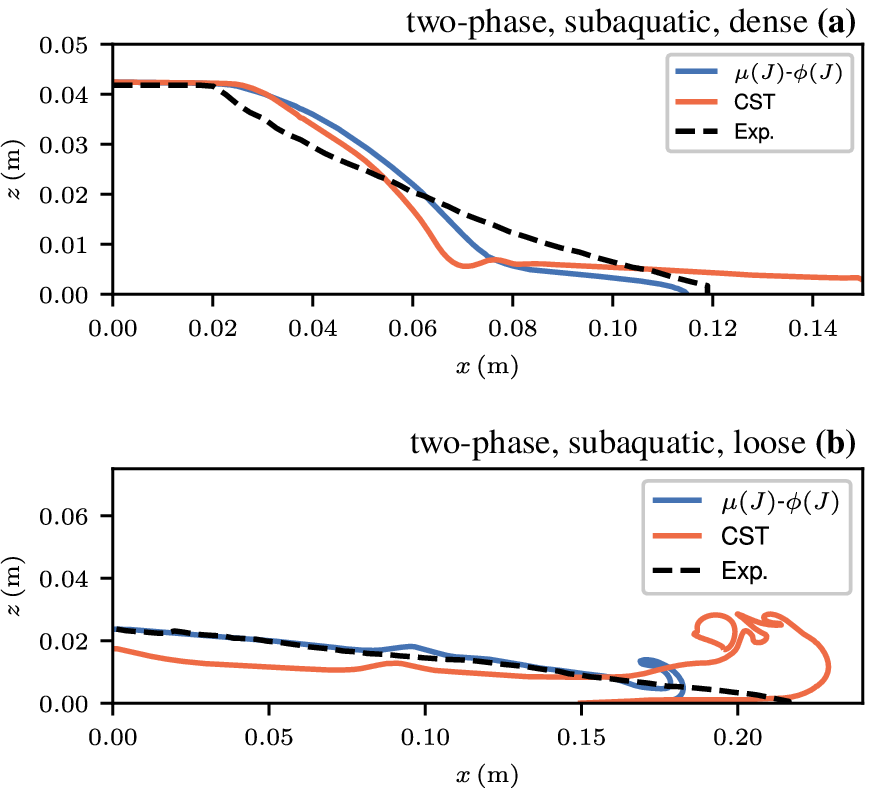}
\end{center}
\caption{The dense granular collapse at $t=6.0\,\r{s}$ (a) and the loose granular collapse at $t=0.65\,\r{s}$ (b), simulated with critical state theory and $\mu(J)$-$\phi(J)$-rheology. The dashed black line shows the final experimental pile shape. The simulations with critical state theory clearly exceed the experiment early in the simulation.}
\label{fig:2011_dynamics}
\end{figure}

The dynamic contribution to effective pressure and friction is imperative for simulations at low Stokes numbers.
Fig.~\ref{fig:2011_dynamics} shows the dense and loose subaquatic two-phase simulations with critical state theory and $\mu(J)$-$\phi(J)$-theory.
The outrun cannot be controlled without the dynamic contributions of the $\mu(J)$-$\phi(J)$-theory and exceeds the final runout very quickly.

\subsection{Influence of the creep shear rate}
\label{ssec:creep_shear}

\begin{figure}
\begin{center}
\includegraphics[scale=1]{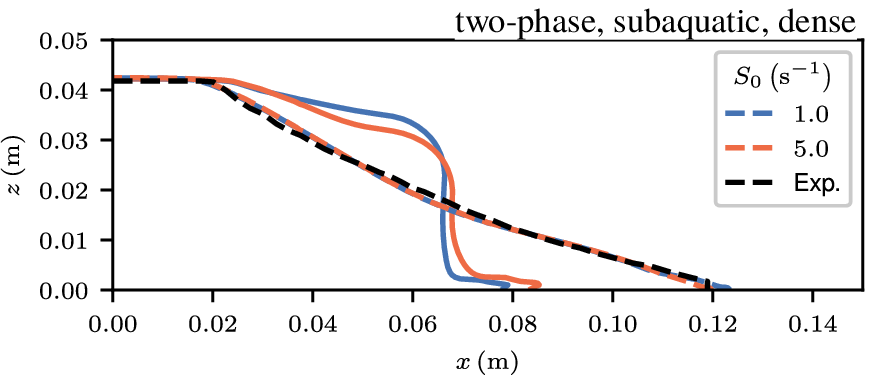}
\end{center}
\caption{Influence of the creep shear rate $S_0$ on the pile shape. Two time steps are shown, $t=1\,\r{s}$ (continuous) and $t=10\,\r{s}$ (dashed). The black dashed line shows the final experimental pile shape.}
\label{fig:2011_s0}
\end{figure}

The influence of the creep shear rate $S_0$ is shown in Fig.~\ref{fig:2011_s0} for the subaquatic dense case.
The Figure shows an early time step at $t=1.0\,\r{s}$ and the final pile shape at $t=10\,\r{s}$.
A reduction of the creeping shear rate from $5\,\r{s^{-1}}$ to $1\,\r{s^{-1}}$ leads to a slower initial collapse of the column.
The delayed collapse is desirable and brings the simulation closer to the experiment.
However, the low value for $S_0$ leads to oscillations in the particle pressure because the simulation accelerates and reaches shear rates beyond $S_0$ too quickly.
The simulation with $S_0 = 5\,\r{s^{-1}}$ shows no oscillations and has thus been utilized in the main part of this work.
Notably, a smaller time step duration will allow smaller values for $S_0$, however for an increased computational cost.
The final pile shape is barley affected by the change in $S_0$.

\bibliographystyle{jfm}


\end{document}